\documentclass[aps,prx,reprint,superscriptaddress,nofootinbib,longbibliography]{revtex4-2}
\usepackage{amsmath,amssymb,amsthm,mathtools}
\usepackage[T1]{fontenc}
\usepackage{lmodern}
\usepackage{physics}
\usepackage{quantikz}
\usepackage{graphicx,graphics,booktabs}
\usepackage[colorlinks=true, citecolor=blue]{hyperref}
\hypersetup{
pdftitle={Rational quantum spectral filtering by repeat-until-success protocol}, 
pdfauthor={M. A. Gavreev, A. K. Fedorov, E. O. Kiktenko}}
\newtheorem{theorem}{Theorem}
\newtheorem{proposition}[theorem]{Proposition}

\newtheorem{definition}[theorem]{Definition}

\begin{document}
\title{Rational quantum spectral filtering by repeat-until-success protocol}

\author{Maksim~A.~Gavreev}
\email{m.gavreev@rqc.ru}
\affiliation{Russian Quantum Center, Skolkovo, Moscow 143025, Russia}
\affiliation{National University of Science and Technology MISIS, Moscow 119049, Russia}

\author{Aleksey~K.~Fedorov}
\email{akf@rqc.ru}
\affiliation{Russian Quantum Center, Skolkovo, Moscow 143025, Russia}
\affiliation{National University of Science and Technology MISIS, Moscow 119049, Russia}

\author{Evgeniy~O.~Kiktenko}
\email{e.kiktenko@rqc.ru}
\affiliation{Russian Quantum Center, Skolkovo, Moscow 143025, Russia}
\affiliation{National University of Science and Technology MISIS, Moscow 119049, Russia}
\date{\today}

\begin{abstract}
We introduce a finite-memory repeat-until-success (RUS) framework for implementing rational spectral filters on a quantum computer, extending spectral filtering beyond the polynomial transformations implemented by quantum signal processing (QSP) and quantum singular value transformation (QSVT).
Our approach interleaves queries to a block encoding of a Hermitian operator $X$, whose spectrum lies in $[-1,1]$, and mid-circuit measurements that either accept, reject, or continue the computation without resetting the quantum state.
Although the protocol may run for arbitrarily many rounds, its terminal statistics are determined by a finite-dimensional Lyapunov equation.
Under uniform termination and phase-inversion symmetry, we characterize exactly the spectral filters $R(X)$ realizable upon acceptance: $R$ can be any real rational function with no poles on $[-1,1]$ and values in $[0,1]$.
We provide a constructive compiler for any admissible $R$, with bounds on memory and expected query cost, and characterize the accepted quantum operation through a spectral coherence kernel.
Applications to regularized resolvents and Lorentzian spectroscopy yield exact low-degree realizations and query-cost advantages over the fixed-depth QSVT polynomial-postselection baseline considered here, with all repetitions and independent restarts after rejection included in the cost.
\end{abstract}
\maketitle

\section{Introduction}
\label{sec:intro}

Quantum computation exploits superposition, interference, and entanglement to approach problems beyond the practical reach of classical computation~\cite{Preskill2018,Montanaro2016, fedorov2022quantum}. 
From quantum simulation to factoring, search, and linear systems~\cite{Feynman1982,Lloyd1996,Shor1997,Grover1996,HarrowEtAl2009,ChildsKothariSomma2017,Szegedy2004}, many quantum algorithms admit a common spectral interpretation: encode an operator and manipulate or probe its eigenvalues or singular values. 
Block-encoding and qubitization provide the corresponding signal model, while quantum signal processing (QSP) and quantum singular-value transformation (QSVT) compile programmable spectral responses into quantum circuits~\cite{BerryEtAl2015,LowChuang2017,LowChuang2019,GilyenEtAl2019,MartynEtAl2021}. 
This approach underlies Hamiltonian simulation, amplitude amplification, eigenvalue filtering, and ground-state algorithms~\cite{YoderEtAl2014,PoulinWocjan2009,LinTong2020,DongLinTong2022,SapovaFedorov2022}, supported by increasingly sophisticated synthesis and generalization techniques~\cite{Haah2019,DongEtAl2021,ChaoEtAl2020,Ying2022,NiYing2026,MotlaghWiebe2024,LuLiuLin2026}.

Finite coherent QSP/QSVT sequences natively implement polynomial or Laurent-polynomial transformations; more general responses are typically accessed through polynomial approximation. 
Other extensions of QSP and QSVT also go beyond polynomial transformations, but mainly through coherent quantum operations~\cite{DongEtAl2024Infinite,AlexisEtAl2026,AlexisMnatsakanyanThiele2024}, while rational transformations based on linear combinations of unitaries (LCU) express resolvents through Hamiltonian-simulation integrals realized by discretization or continuous-variable resources~\cite{ShenEtAl2024Rational}. 
Rational approximation is also known to be closely related to postselected quantum query algorithms~\cite{MahadevDeWolf2015}, but that correspondence does not provide a universal circuit realization of rational spectral effects. 
We instead ask whether a finite-memory architecture with mid-circuit measurements can implement such effects \emph{exactly}, without first approximating the desired rational function by a polynomial. 
Rational functions are a compelling target: their denominators efficiently encode sharp spectral features and characteristic scales, as established in rational approximation theory~\cite{Akhiezer1956,Newman1964,TrefethenWilber2025,Gawlik2019}. 
They are also central to matrix functions, rational Krylov methods, contour eigensolvers, and adaptive numerical approximation~\cite{Higham2008,HaleHighamTrefethen2008,Guettel2013,Polizzi2009,GuettelPolizziTangViaud2015,NakatsukasaEtAl2018,AlaouiMahoney2015,MuscoMusco2017}. 
Resolvents, regularized inverses, Green functions, and Lorentzian windows are intrinsically rational~\cite{Economou2006,TrefethenWeideman2014}. 
Measurement feedforward and recursive eigenvalue transformations further motivate broader spectral processing~\cite{DongAnNiu2024,MizutaFujii2024}.

\begin{figure*}[ht!]
    \centering
    \includegraphics[width=1.\textwidth]{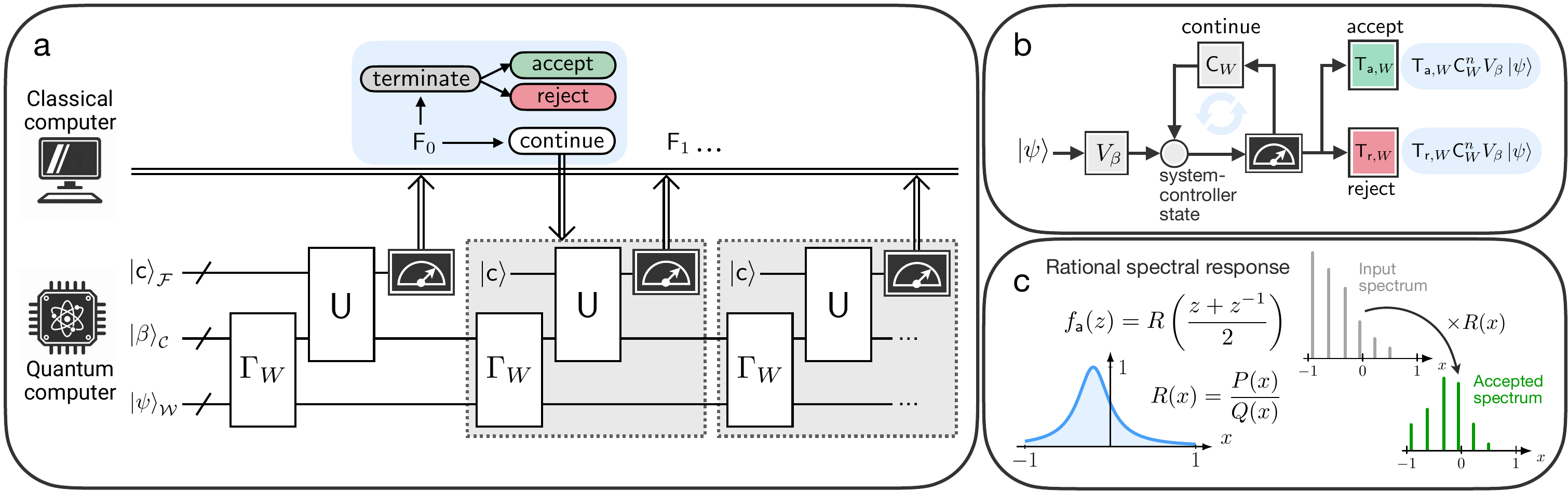}
    \vspace{-12pt}
    \caption{Rational quantum spectral filtering by RUS. (a) Protocol architecture. A controlled-walk router $\Gamma_W$ and a fixed unitary colligation $\mathsf{U}$ act on the signal, controller memory, and measurement flag. Each round produces one of three outcomes. (b) First-hitting representation of the protocol. The continuation operator $C$ generates successive controller states, while terminal operators $\mathsf{T}_{\mathsf{a}}$ and $\mathsf{T}_{\mathsf{r}}$ describe acceptance and rejection after an arbitrary number of iterations. (c) Rational spectral response. Summation over all first-hitting histories produces a rational acceptance probability $R(x)=P(x)/Q(x)$, which reweights the input spectral distribution and enables selective spectral filtering.}
    \label{fig:main}
\end{figure*}

Here we introduce rational quantum spectral filtering through measured repeat-until-success (RUS) dynamics, building on conditional-repetition protocols developed for gate synthesis~\cite{PaetznickSvore2014,BocharovEtAl2015,SoedaMurao2021}. 
A qubitized walk interacts with a finite controller through a controlled query and a fixed unitary colligation. 
Each round is measured as continue, accept, or reject; continuation retains the conditional signal--memory state and repeats the same cell. 
The same circuit can run for any number of rounds. The final acceptance probability is the sum over all possible stopping times.
Since a measurement is performed after each round, different stopping times contribute probabilities rather than coherently added amplitudes~\cite{DaviesLewis1970,WisemanMilburn2009,Watrous2018}, producing a rational acceptance response.
The denominator is generated by the measured first-hitting recurrence itself, so the eventual acceptance effect is exactly rational in the ideal uniformly terminating protocol. 
Experimental implementations of QSP and continuous device monitoring provide additional context~\cite{KikuchiEtAl2023,ZolotarevEtAl2023}.

The mechanism is naturally described by classical realization and systems theory~\cite{Kalman1963,Kailath1980,ZhouDoyleGlover1996}. 
Related finite-memory dynamical descriptions appear in data-driven studies of non-Markovian quantum dynamics~\cite{LuchnikovEtAl2022}. 
For each walk eigenphase, the controller follows a finite-dimensional recurrence. The sum over all stopping times satisfies a discrete Lyapunov equation.
Its finite resolvent is rational in the phase; phase-inversion symmetry converts it into a function of the Hermitian spectrum. 
This mirrors the finite- versus infinite-impulse-response (FIR/IIR) distinction in digital signal processing~\cite{Vaidyanathan1993,Kailath1980}. 
Importantly, the analogy concerns the spectral response and feedback structure: distinct RUS stopping histories contribute probabilities, not coherently summed transfer-function amplitudes. 
Unitary colligations, spectral factorization, and lossless factorizations supply constructive tools~\cite{Schur1917,WienerMasani1957,Vaidyanathan1993,Dumitrescu2007,Alpay2001,BallBolotnikov2004,DritschelRovnyak2010}.

Our central result is an exact representation theorem: the universally realizable, uniformly terminating, phase-symmetric acceptance effects are precisely the real rational functions without poles on the signal interval and bounded between zero and one. 
Both directions are established. 
For the converse direction, we give a constructive compiler based on polynomial trial probabilities, spectral factorization, and controlled-walk operations~\cite{Fejer1916,Riesz1916,DritschelRovnyak2010,BerntsonSunderhauf2025}.
We derive quantitative rational-degree, memory, termination, and expected-query bounds, including freedom to optimize the renewal stopping profile.

We also characterize the complete terminal quantum operation, not only its acceptance effect. 
A positive-definite two-variable rational coherence kernel, obtained from a mixed Stein equation, determines spectral coherences while its diagonal fixes acceptance probabilities. 
For canonical block-encoding, accepted runs exactly reweight the input spectral distribution. 
Applications include an exact one-query regularized-resolvent filter for low-energy weighting and sampling, and a degree-two Lorentzian filter for broadened spectroscopy and Green-function observables, connecting to quantum spectral-density and response-function algorithms~\cite{Roggero2020,EfrosEtAl2007,Rall2020}. 
End-to-end comparisons with fixed-depth polynomial QSP/QSVT demonstrate expected-query savings in the specified fresh-copy, terminal-postselection model.

The paper is organized as follows.
Section~\ref{sec:rus} introduces the protocol, and Sec.~\ref{sec:response} establishes the representation theorem.
Section~\ref{sec:compiler} presents the constructive compiler and derives resource bounds, while Sec.~\ref{sec:applications} explores applications.
Section~\ref{sec:discussion} concludes with a discussion of the results and future directions.
Technical proofs are provided in the appendices.

\section{Spectral filtering by RUS protocol}
\label{sec:rus}

We begin with a general overview of the RUS protocol and its filtering action, leaving precise definitions and technical details to the subsequent sections.

Our protocol, illustrated in Fig.~\ref{fig:main}(a), uses three quantum registers: a walk register $\mathcal{W}$ carrying the state to be filtered, a finite-dimensional controller memory $\mathcal{C}$ retaining quantum information between rounds, and an auxiliary measurement flag $\mathcal{F}$.
Each round applies the same finite quantum circuit, followed by a measurement of the flag with outcomes $\mathsf{accept}$, $\mathsf{reject}$, or $\mathsf{continue}$.
Acceptance or rejection terminates the protocol.
Upon continuation, the conditional joint state of $\mathcal{W}$ and $\mathcal{C}$ is retained for the next round, while only the flag is reset.
A fixed circuit with finite quantum memory can therefore be executed within arbitrarily many rounds, with classical feedforward determining whether to continue or stop.
The protocol thus follows the logic of a classical do--while loop. It executes a round before checking whether to repeat, and terminates with one of the two conditional output states associated with the $\mathsf{accept}$ and $\mathsf{reject}$ outcomes, as illustrated in Fig.~\ref{fig:main}(b).

Of the two possible terminal outcomes, we focus on $\mathsf{accept}$. To relate the corresponding output state to the input, we consider how the input and output states of the walk register are distributed over the eigenspaces of a Hermitian operator $X$. We refer to these distributions as the input and output spectra, respectively.
The operator $X$ enters the circuit through a unitary $W$ that block-encodes it. In each round, a controlled application of $W$, denoted by $\Gamma_W$, couples the walk register to the controller memory.

The protocol acts as a spectral filter because the output spectrum is obtained by multiplying the input weight at each eigenvalue $x$ by $R(x)$ and renormalizing the resulting weights [see Fig.~\ref{fig:main}(c)]. As we show below, $R$, which we call the spectral response of the filter, is a rational function of $x$ and can be written as a ratio of two real polynomials. Importantly, the form of $R$ is independent of $X$. We next describe the protocol in more detail.

\subsection{Signal encoding}
\label{sec:signal}

We begin by specifying how the Hermitian operator that defines the filtering task enters the circuit.
Let $H$ be a Hermitian operator acting on a finite-dimensional system space $\mathcal{H}_{\mathcal{S}}$.
Given a known scale $\alpha>0$ satisfying $\|H\|\leq\alpha$, where $\|\cdot\|$ denotes the operator norm, we define
\begin{equation}
X=\frac{H}{\alpha},
\qquad
\operatorname{spec}(X)\subseteq[-1,1].
\label{eq:normalize}
\end{equation}
Here, $\operatorname{spec}(X)$ denotes the set of eigenvalues of $X$.
Alternatively, a known spectral interval can be mapped to $[-1,1]$ by an affine rescaling.
The spectral decomposition is
\begin{equation}
X=\sum_\lambda x_\lambda P_\lambda,
\label{eq:spectral-decomp}
\end{equation}
Here $\lambda$ labels the distinct eigenvalues $x_\lambda$, and $P_\lambda$ denotes the orthogonal projector onto the corresponding eigenspace.

We introduce an auxiliary qubit $\mathcal{B}$ and define the extended space
$\mathcal{H}_{\mathcal{W}}=\mathcal{H}_{\mathcal{B}}\otimes\mathcal{H}_{\mathcal{S}}$,
which we call the walk space.
We assume access to controlled applications of the unitary operator
\begin{equation}
W=
\begin{pmatrix}
X & \sqrt{I_{\mathcal{S}}-X^2}\\
-\sqrt{I_{\mathcal{S}}-X^2} & X
\end{pmatrix},
\label{eq:walk}
\end{equation}
where the blocks are written in the computational basis of $\mathcal{B}$. Throughout, $I_{\mathcal{A}}$ denotes the identity operator on the space $\mathcal{A}$.
We refer to $W$ as a quantum walk, with each application representing one discrete step of unitary evolution on the signal register.
We note that $W$ can be obtained from the canonical block encoding of $X$ with off-diagonal blocks $i\sqrt{I_{\mathcal{S}}-X^2}$ by conjugation with the single-qubit phase gate $\operatorname{diag}(1,i)$ acting on $\mathcal{B}$.

To understand this evolution, consider the tensor product of the auxiliary-qubit space $\mathcal{H}_{\mathcal{B}}$ and the eigensubspace of $X$ corresponding to $x_\lambda$.
Within this subspace, $W$ acts as
\begin{equation}
W_{\lambda}=
\begin{pmatrix}
x_{\lambda} & \sqrt{1-x_{\lambda}^2}\\
-\sqrt{1-x_{\lambda}^2} & x_{\lambda}
\end{pmatrix}\otimes I_{{\rm rank}(P_{\lambda})}.
\end{equation}
Writing $\theta_{\lambda}=\arccos x_{\lambda}$, the operator acting on the auxiliary qubit takes the form $e^{i\theta_{\lambda}\sigma_y}$, where $\sigma_y$ is the Pauli $y$ matrix.

The eigenstates of $W$ can therefore be written as
\begin{equation}
\ket{\omega_\pm(x)}
:=\ket{v_\pm}_{\mathcal{B}}\otimes\ket{x}_{\mathcal{S}},
\end{equation}
where $\ket{v_\pm}:=2^{-1/2}(\ket{0}\pm i\ket{1})$ and $\ket{x}$ is an eigenvector of $X$ with eigenvalue $x\in\{x_\lambda\}$.
The corresponding eigenvalues of $W$ are
\begin{equation}
z_\pm(x)=x\pm i\sqrt{1-x^2}
=e^{\pm i\arccos x}.
\label{eq:phase-map}
\end{equation}
So we see that each system eigenvector gives rise to two eigenstates of $W$, distinguished by the auxiliary-qubit state.
At $x=\pm1$, the two eigenvalues coincide, while the two states remain orthogonal.
Conversely, $x$ can be recovered from either corresponding eigenvalue $z$ of $W$:
\begin{equation}
x=\frac{z+z^{-1}}{2},
\qquad z\in\{z_+(x),z_-(x)\}.
\label{eq:inverse-phase-map}
\end{equation}

For convenience, we denote an eigenstate of $W$ with eigenvalue $z$ by $\ket{\omega_z}$, so that
\begin{equation}
W\ket{\omega_z}=z\ket{\omega_z}.
\end{equation}
If the eigenvalue is degenerate, $\ket{\omega_z}$ denotes any state in the corresponding eigenspace.
In what follows, we refer to the eigenspace of $W$ corresponding to an eigenvalue $z$ as the phase sector associated with $z$.

As the initial state of the walk register, we take $\ket{0}_{\mathcal{B}}\otimes\ket{\psi}_{\mathcal{S}}$ for an arbitrary system state $\ket{\psi}_{\mathcal{S}}$. This state is obtained via the isometry
\begin{equation}
\mathcal{E}\ket{\psi}_{\mathcal{S}}
=\ket{0}_{\mathcal{B}}\otimes\ket{\psi}_{\mathcal{S}},
\label{eq:signal-embedding}
\end{equation}
which we call the signal embedding.

For later use, we denote the Hermitian part of the walk operator by
\begin{equation} \label{eq:X_W}
X_{\mathcal{W}}
:=\frac{W+W^\dagger}{2}
=I_{\mathcal{B}}\otimes X.
\end{equation}
Since $X_{\mathcal{W}}$ acts as $X$ on the system and leaves the auxiliary qubit unchanged, applying $X_{\mathcal{W}}$ after the embedding is equivalent to applying $X$ before it:
\begin{equation}
X_{\mathcal{W}}\mathcal{E}=\mathcal{E}X.
\label{eq:bridge}
\end{equation}

\subsection{Controlled-walk and routing}
\label{sec:walk}
In our protocol, we control the application of $W$ using a finite-dimensional register $\mathcal{C}$ with $\mathcal{H}_{\mathcal{C}}\simeq\mathbb{C}^d$, called the controller memory.
Let $\Pi_s$ be an orthogonal projector onto a subspace of $\mathcal{H}_{\mathcal{C}}$, which we call the active signal route.
We couple the controller memory $\mathcal{C}$ to the walk register $\mathcal{W}$ through the controlled-$W$ operation
\begin{equation}
\Gamma_W = (I_{\mathcal{C}}-\Pi_s)\otimes I_{\mathcal{W}} +\Pi_s\otimes W,
\label{eq:router}
\end{equation}
which we call the signal router.
This is the only signal-dependent operation in each elementary round of the protocol.

For a walk eigenstate $\ket{\omega_z}$, the signal router acts as
\begin{equation}
\Gamma_W(\ket{\xi}\otimes\ket{\omega_z})
=\bigl(\Gamma(z)\ket{\xi}\bigr)\otimes\ket{\omega_z}
\end{equation}
with
\begin{equation}\label{eq:phase-router}
\Gamma(z)=I_{\mathcal{C}}-\Pi_s+z\Pi_s,
\end{equation}
for any $\ket{\xi}\in\mathcal{H}_{\mathcal{C}}$.
Thus, the router applies the unitary $\Gamma(z)$ to the controller memory while leaving the walk eigenstate unchanged.

Notably, the router applies $W$ without requiring $W^\dagger$.
Consequently, any finite sequence
\begin{equation}
V_L\Gamma(z)V_{L-1}\cdots V_1\Gamma(z)V_0,
\label{eq:causal-sequence}
\end{equation}
with signal-independent unitaries $V_\ell$ has matrix elements that are polynomials in $z$ of degree at most $L$, containing only nonnegative powers.
Negative powers can appear when forming Born probabilities, because complex conjugation replaces $z$ with $z^{-1}$ on the unit circle.
In Sec.~\ref{sec:compiler}, we use this polynomial structure together with Fej\'er--Riesz factorization to construct the required trial amplitudes.

\subsection{Unitary colligation}
\label{sec:colligation}

The remaining operation in each round is the application of a fixed, signal-independent unitary $\mathsf{U}$ to the controller memory $\mathcal{C}$ and a three-dimensional flag register $\mathcal{F}$, with
\[
\mathcal{H}_{\mathcal{F}}
=\operatorname{span}\{
\ket{\mathsf{c}},\ket{\mathsf{a}},\ket{\mathsf{r}}\}.
\]
This is followed by a measurement of $\mathcal{F}$ in this basis, with outcomes corresponding to continuation, acceptance, and rejection, respectively.
We refer to $\mathsf{U}$, together with its flag-block structure, as a unitary colligation.
In the ordered flag basis $(\ket{\mathsf{c}},\ket{\mathsf{a}},\ket{\mathsf{r}})$, it takes the form
\begin{equation}
\mathsf{U}
=\left(
\begin{array}{c|cc}
\mathsf{C} & * & *\\
\hline
\mathsf{T}_{\mathsf{a}} & * & *\\
\mathsf{T}_{\mathsf{r}} & * & *
\end{array}
\right),
\label{eq:colligation}
\end{equation}
where each block acts on $\mathcal{H}_{\mathcal{C}}$.
The asterisks denote blocks that do not affect a round with this flag initialization.

For any controller state $\ket{\xi}\in\mathcal{H}_{\mathcal{C}}$ and the flag initialized in $\ket{\mathsf{c}}$, the colligation acts as
\begin{equation}
\mathsf{U}\bigl(\ket{\mathsf{c}}\otimes\ket{\xi}\bigr)
=\ket{\mathsf{c}}\otimes\mathsf{C}\ket{\xi}+\ket{\mathsf{a}}\otimes\mathsf{T}_{\mathsf{a}}\ket{\xi}+\ket{\mathsf{r}}\otimes\mathsf{T}_{\mathsf{r}}\ket{\xi}.
\label{eq:colligation-action}
\end{equation}
The unitarity of $\mathsf{U}$ implies that its first block column is an isometry:
\begin{equation}
\mathsf{C}^\dagger\mathsf{C}
+\mathsf{T}_{\mathsf{a}}^\dagger\mathsf{T}_{\mathsf{a}}
+\mathsf{T}_{\mathsf{r}}^\dagger\mathsf{T}_{\mathsf{r}}
=I_{\mathcal{C}}.
\label{eq:first-column-isometry}
\end{equation}

With the flag initialized in $\ket{\mathsf{c}}$, one round consists of applying the router $\Gamma_W$, followed by the unitary $\mathsf{U}$ and a projective measurement of the flag.
The corresponding branch operators on the remaining registers,
$\mathcal{H}_{\mathcal{C}}\otimes\mathcal{H}_{\mathcal{W}}$, are
\begin{equation}
\begin{aligned}
\mathsf{C}_W
&=(\mathsf{C}\otimes I_{\mathcal{W}})\Gamma_W,\\
\mathsf{T}_{j,W}
&=(\mathsf{T}_j\otimes I_{\mathcal{W}})\Gamma_W,
\quad j\in\{\mathsf{a},\mathsf{r}\}.
\end{aligned}
\label{eq:full-branches}
\end{equation}
These are the Kraus operators associated with the continuation, acceptance, and rejection outcomes, respectively.
They determine the unnormalized conditional states of the controller memory and the walk register and satisfy the completeness relation
\begin{equation}
\mathsf{C}_W^\dagger\mathsf{C}_W
+\sum_{j\in\{\mathsf{a},\mathsf{r}\}}
\mathsf{T}_{j,W}^\dagger\mathsf{T}_{j,W}
=I_{\mathcal{C}}\otimes I_{\mathcal{W}}.
\label{eq:one-round-complete-full}
\end{equation}

Since the router preserves each eigenspace of $W$, it is convenient to introduce the controller matrices
\begin{equation}
\begin{aligned}
\mathsf{C}(z)&=\mathsf{C}\Gamma(z),\\
\mathsf{T}_j(z)&=\mathsf{T}_j\Gamma(z),
\quad j\in\{\mathsf{a},\mathsf{r}\},
\end{aligned}
\label{eq:phase-branches}
\end{equation}
which determine the action of the branch operators within the eigenspace corresponding to $z$.
These matrices describe the conditional evolution of the controller memory, while the walk state remains unchanged.

\subsection{First-hitting histories and termination}
\label{sec:histories}

We now describe the measurement histories generated by repeated rounds of the protocol and give a condition ensuring eventual termination. This condition also provides bounds on the probability of continuing beyond a given number of rounds and on the expected stopping time.
We initialize the controller memory in a normalized state $\ket{\beta}\in\mathcal{H}_{\mathcal{C}}$ and define the isometry
$V\beta\ket{\psi}=\ket{\beta}\otimes\ket{\psi}$.
A measurement history consisting of $n$ consecutive continuation outcomes followed by a terminal outcome $j\in\{\mathsf{a},\mathsf{r}\}$ is described by the first-hitting Kraus operator
\begin{equation}
K_{j,n}=\mathsf{T}_{j,W}\mathsf{C}_W^n V_\beta,
\quad n=0,1,2,\ldots
\label{eq:first-hitting-operator}
\end{equation}
The factor $\mathsf{C}^n_W$ represents $n$ continuing rounds, and $\mathsf{T}_{j,W}$ gives the terminal outcome.
Different stopping times correspond to mutually exclusive measurement records and therefore contribute incoherently to the terminal statistics.

Since both the router and colligation preserve the eigenspaces of $W$, every history admits a finite-dimensional spectral reduction.
For a walk eigenstate $W\ket{\omega_z}=z\ket{\omega_z}$, we have
\begin{equation}
\begin{aligned}
K_{j,n}\ket{\omega_z}
&=\ket{k_{j,n}(z)}\otimes\ket{\omega_z},\\
\ket{k_{j,n}(z)}
&=\mathsf{T}_j(z)\mathsf{C}(z)^n\ket{\beta}.
\end{aligned}
\label{eq:first-hit-vector}
\end{equation}
Thus, the evolution within each walk-phase sector is completely determined by $d$-dimensional matrices acting on the controller memory.

To ensure termination uniformly over all phase sectors, we assume that there exist an integer $m\geq1$ and a constant $0<\eta<1$ such that
\begin{equation}
\sup_{z\in\mathbb{T}}\|\mathsf{C}(z)^m\|\leq\eta,
\label{eq:termination}
\end{equation}
where $\mathbb{T}:=\{z\in\mathbb{C}:|z|=1\}$.
Let $T$ denote the number of measured rounds up to and including the terminal outcome.
Since $\|\mathsf{C}(z)\|\leq1$, the probability of continuing beyond $n$ rounds satisfies
\begin{equation}
\Pr(T>n\mid z)
=\|\mathsf{C}(z)^n\ket{\beta}\|^2
\leq\eta^{2\lfloor n/m\rfloor}.
\label{eq:stopping-tail-general}
\end{equation}
The protocol therefore terminates with probability one.
A simple sufficient condition for Eq.~\eqref{eq:termination} is $\|\mathsf{C}\|<1$: since $\Gamma(z)$ is unitary, $\|\mathsf{C}(z)\|=\|\mathsf{C}\|$, so one may take $m=1$ and any $0<\eta<1$ satisfying $\|\mathsf{C}\|\leq\eta$.

The tail-sum identity gives the expected stopping time as
\begin{equation}
\mathbb{E}[T\mid z]
=\sum_{n=0}^{\infty}\Pr(T>n\mid z)
=\bra{\beta}Z(z)\ket{\beta}
\leq\frac{m}{1-\eta^2},
\label{eq:expected-stopping-general}
\end{equation}
where
\begin{equation}
Z(z)=\sum_{n=0}^{\infty}
\bigl(\mathsf{C}(z)^\dagger\bigr)^n\mathsf{C}(z)^n.
\label{eq:stopping-gramian}
\end{equation}
This series converges uniformly in $z$, and $Z(z)$ is the unique positive-definite solution of the discrete Lyapunov equation
\begin{equation}
Z(z)=I_{\mathcal{C}}+\mathsf{C}(z)^\dagger Z(z)\mathsf{C}(z).
\label{eq:stopping-lyapunov}
\end{equation}
Thus, the expected stopping time can be computed from a finite-dimensional linear equation and is bounded uniformly over all phase sectors.

The one-round completeness relation can be iterated over successive measurement histories. 
Summing all terminal events within the first $N$ rounds gives
\begin{equation}
\begin{aligned}
\sum_{j\in\{\mathsf{a},\mathsf{r}\}}
\sum_{n=0}^{N-1}
K_{j,n}^\dagger K_{j,n}+ V_\beta^\dagger
(\mathsf{C}_W^\dagger)^N
\mathsf{C}_W^N V_\beta
=I_{\mathcal{W}}.
\end{aligned}
\label{eq:finite-normalization-main}
\end{equation}
The second term represents the unresolved continuation branch after $N$ rounds.
Under uniform termination it vanishes as $N\to\infty$, so the terminal outcomes exhaust the total probability.

This construction has a natural interpretation in linear systems theory: the retained controller memory acts as a finite internal state, while $\mathsf{C}(z)$ governs its discrete-time evolution.
The stability condition ensures that the infinite sequence of conditional operations is well defined.
We now use this structure to characterize the resulting quantum instrument.

\subsection{Terminal operations and POVM representation}
\label{sec:terminal}

We now describe the terminal outcome probabilities and the corresponding output states, which together define a quantum instrument.
For an input state $\rho_{\mathcal{W}}$ of the walk register $\mathcal{W}$, comprising the auxiliary qubit $\mathcal{B}$ and the system register $\mathcal{S}$, the operation associated with the terminal outcome $j\in\{\mathsf{a},\mathsf{r}\}$ is
\begin{equation}
\Phi_j(\rho_{\mathcal{W}})
=\sum_{n=0}^{\infty}
\operatorname{Tr}_{\mathcal{C}}
\left[K_{j,n}\rho_{\mathcal{W}}K_{j,n}^\dagger\right].
\label{eq:terminal-operation}
\end{equation}
Here, we discard the controller memory and retain only the terminal outcome, summing over all stopping times.

Each $\Phi_j$ is completely positive and trace-nonincreasing.
The corresponding POVM effect is
\begin{equation}
F_j=\Phi_j^*(I_{\mathcal{W}})
=\sum_{n=0}^{\infty}K_{j,n}^\dagger K_{j,n},
\label{eq:terminal-effect}
\end{equation}
where $\Phi_j^*$ denotes the adjoint map with respect to the Hilbert--Schmidt inner product.
The probability of outcome $j$ is therefore
\begin{equation}
p_j=\operatorname{Tr}(F_j\rho_{\mathcal{W}}).
\label{eq:terminal-probability-state}
\end{equation}
Uniform termination implies
\begin{equation}
F_{\mathsf{a}}+F_{\mathsf{r}}=I_{\mathcal{W}},
\qquad
0\preceq F_j\preceq I_{\mathcal{W}}.
\label{eq:terminal-povm-completeness}
\end{equation}
Thus, $\{F_{\mathsf{a}},F_{\mathsf{r}}\}$ is a complete POVM, and $\Phi_{\mathsf{a}}+\Phi_{\mathsf{r}}$ is a quantum channel.

To characterize the action of the instrument, consider the spectral decomposition of the walk operator,
\begin{equation}
W=\sum_\nu z_\nu\Omega_\nu,
\end{equation}
where $z_\nu$ are the distinct eigenvalues of $W$ and $\Omega_\nu$ are the orthogonal projectors onto the corresponding eigenspaces (recall that these eigenvalues occur in complex-conjugate pairs according to Eq.~\eqref{eq:phase-map}, with the two members coinciding at $x=\pm1$).

Within each phase sector, a first-hitting history leaves the walk state unchanged and produces a controller vector $\ket{k_{j,n}(z)}$ defined in Eq.~\eqref{eq:first-hit-vector}.
Consequently,
\begin{equation} \label{eq:K_jn}
K_{j,n}
=\sum_\nu\ket{k_{j,n}(z_\nu)}\otimes\Omega_\nu.
\end{equation}
Using the orthogonality of the spectral projectors, we obtain
\begin{equation}
K_{j,n}^\dagger K_{j,n}
=\sum_\nu\|k_{j,n}(z_\nu)\|^2\Omega_\nu.
\end{equation}
The terminal effect therefore takes the spectral form
\begin{equation}
F_j=\sum_\nu f_j(z_\nu)\Omega_\nu,
\label{eq:phase-effect}
\end{equation}
where
\begin{equation}
f_j(z)=\sum_{n=0}^{\infty}
\left\|\mathsf{T}_j(z)\mathsf{C}(z)^n\ket{\beta}\right\|^2.
\label{eq:phase-response}
\end{equation}
The scalar response $f_j(z)$ is the eventual probability of outcome $j$ for any normalized input state within the phase sector $z$.

The POVM effects determine the terminal outcome probabilities, while the quantum operations $\Phi_j$ also describe how coherences between phase sectors are transformed.
Substituting the spectral form of the first-hitting operators~\eqref{eq:K_jn} into Eq.~\eqref{eq:terminal-operation} yields
\begin{equation}
\Phi_j(\rho_{\mathcal{W}})
=\sum_{\nu,\mu}\chi_j(z_\nu,z_\mu)
\Omega_\nu\rho_{\mathcal{W}}\Omega_\mu,
\label{eq:walk-channel}
\end{equation}
where
\begin{equation}
\begin{aligned}
\chi_j(z,w)
&=\sum_{n=0}^{\infty}
\langle k_{j,n}(w)|k_{j,n}(z)\rangle\\
&=\sum_{n=0}^{\infty}
\bra{\beta}
\bigl(\mathsf{C}(w)^\dagger\bigr)^n
\mathsf{T}_j(w)^\dagger
\mathsf{T}_j(z)
\mathsf{C}(z)^n
\ket{\beta}
\end{aligned}
\label{eq:chi}
\end{equation}
is the spectral coherence kernel.
This kernel completely specifies the terminal operation in the spectral representation of $W$.
Its diagonal gives the POVM response,
\begin{equation}
\chi_j(z,z)=f_j(z),
\label{eq:chi-diagonal}
\end{equation}
while its off-diagonal values determine how coherences between distinct phase sectors are transformed.

To make the structure of the coherence kernel explicit, we collect the controller vectors associated with all stopping times into a single history vector,
\begin{equation}
\ket{h_j(z)}
=\bigoplus_{n=0}^{\infty}\ket{k_{j,n}(z)}.
\label{eq:history-feature}
\end{equation}
Different stopping times occupy orthogonal components of this direct sum, reflecting their distinct measurement records.
Since $\|h_j(z)\|^2=f_j(z)\leq1$, these vectors are well defined, and the kernel can be written as
\begin{equation}
\chi_j(z,w)=\langle h_j(w)|h_j(z)\rangle.
\label{eq:chi-gram}
\end{equation}
This representation shows that $\chi_j$ is a positive-semidefinite kernel.
Hermitian symmetry and the Cauchy--Schwarz inequality give
\begin{equation}
\begin{aligned}
\chi_j(z,w)&=\overline{\chi_j(w,z)},\\
|\chi_j(z,w)|^2&\leq f_j(z)f_j(w).
\end{aligned}
\label{eq:chi-properties}
\end{equation}
For $f_j(z)f_j(w)>0$, the normalized overlap
$\chi_j(z,w)/\sqrt{f_j(z)f_j(w)}$
describes the coherence retained between the two phase sectors relative to their respective terminal weights.

We now return to the original system register $\mathcal{S}$ and the Hermitian operator $X=\sum_\lambda x_\lambda P_\lambda$ with which we started.
Recall that the system input is embedded into the walk register via $\mathcal{E}\ket{\psi}=\ket{0}_{\mathcal{B}}\otimes\ket{\psi}_{\mathcal{S}}$.
After the terminal operation, we discard the auxiliary qubit $\mathcal{B}$, obtaining the operation
\begin{equation}
\Psi_j(\rho_{\mathcal{S}})
=\operatorname{Tr}_{\mathcal{B}}
\left[\Phi_j\bigl(\mathcal{E}\rho_{\mathcal{S}}\mathcal{E}^\dagger\bigr)\right],
\label{eq:system-operation}
\end{equation}
which acts solely on the system register $\mathcal{S}$.

To evaluate $\Psi_j$, we use the decomposition $\ket{0}_{\mathcal{B}}=2^{-1/2}(\ket{v_+}_{\mathcal{B}}+\ket{v_-}_{\mathcal{B}})$.
Thus, each system eigenstate with eigenvalue $x$ is embedded into an equal superposition of two walk eigenstates with eigenvalues $z_+(x)$ and $z_-(x)$.
The operation $\Phi_j$ transforms coherences between these walk eigenstates according to the kernel $\chi_j$.
When we trace out $\mathcal{B}$, terms connecting $\ket{v_+}$ and $\ket{v_-}$ vanish because these states are orthogonal.
The resulting system operation therefore averages the contributions from the two branches:
\begin{equation}
\Psi_j(\rho_{\mathcal{S}})
=\sum_{\lambda,\kappa}
\Xi_j(x_\lambda,x_\kappa)
P_\lambda\rho_{\mathcal{S}}P_\kappa,
\label{eq:system-channel}
\end{equation}
where
\begin{equation}
\Xi_j(x,y)
=\frac{1}{2}\left[
\chi_j\bigl(z_+(x),z_+(y)\bigr)
+\chi_j\bigl(z_-(x),z_-(y)\bigr)
\right].
\label{eq:Xi}
\end{equation}
The system coherence kernel $\Xi_j(x,y)$ specifies how each spectral block $P_\lambda\rho_{\mathcal{S}}P_\kappa$ is transformed.
For $x=y$, it gives the probability of terminal outcome $j$ for an input supported in the eigenspace corresponding to $x$:
\begin{equation}
\Xi_j(x,x)
=\frac{1}{2}
\left[f_j(z_+(x))+f_j(z_-(x))\right].
\label{eq:system-response}
\end{equation}
Consequently, the POVM effect associated with outcome $j$ on the system register is
\begin{equation}
E_j
=\Psi_j^*(I_{\mathcal{S}})
=\mathcal{E}^\dagger F_j\mathcal{E}
=\sum_\lambda\Xi_j(x_\lambda,x_\lambda)P_\lambda.
\label{eq:system-effect}
\end{equation}

For a system input $\rho_{\mathcal{S}}$, the probability of terminal outcome $j$ is
$p_j(\rho_{\mathcal{S}})=\operatorname{Tr}(E_j\rho_{\mathcal{S}})$.
If this probability is nonzero, the corresponding conditional output state is
$\rho_j=\Psi_j(\rho_{\mathcal{S}})/p_j(\rho_{\mathcal{S}})$.
The population of each eigenspace of $X$ is therefore reweighted by the diagonal response $\Xi_j(x_\lambda,x_\lambda)$ and normalized by the total outcome probability:
\begin{equation}
\operatorname{Tr}(P_\lambda\rho_j)
=\frac{
\Xi_j(x_\lambda,x_\lambda)
\operatorname{Tr}(P_\lambda\rho_{\mathcal{S}})
}{
p_j(\rho_{\mathcal{S}})
}.
\label{eq:spectral-law}
\end{equation}
Thus, conditioning on a terminal outcome implements a spectral filter on $X$, with weights determined by the diagonal response $\Xi_j(x,x)$.

\section{Rational response}
\label{sec:response}

The preceding section describes the terminal probabilities and output states as sums over all possible stopping times.
We now establish which spectral filters the protocol can realize and at what resource cost.
First, we show that finite controller memory leads to rational terminal responses through a Lyapunov equation.
We then introduce the symmetry that expresses these responses as functions of the Hermitian operator $X$ and prove an exact characterization of the realizable filters.
Finally, we analyze the memory requirements, termination time, and expected query cost of the constructive realization.

\subsection{Lyapunov equation}
\label{subsec:lyapunov}

Consider a fixed walk phase sector specified by $z\in\mathbb{T}$.
The probability of terminal outcome $j\in\{\mathsf{a},\mathsf{r}\}$, given by Eq.~\eqref{eq:phase-response}, can be written as
$f_j(z)=\bra{\beta}Y_j(z)\ket{\beta}$, where
\begin{equation}
Y_j(z)=\sum_{n=0}^{\infty}
\bigl(\mathsf{C}(z)^\dagger\bigr)^n
S_j(z)\mathsf{C}(z)^n,
\label{eq:lyapunov-series}
\end{equation}
and 
\begin{equation}
    S_j(z)=\mathsf{T}_j(z)^\dagger\mathsf{T}_j(z).
\end{equation}
Uniform stability guarantees that this series converges in operator norm.

Rather than evaluating the infinite sum directly, we can determine $Y_j(z)$ through a finite-dimensional linear equation.
Separating the $n=0$ term and factoring out the continuation matrices from the remaining terms gives
\begin{equation}
Y_j(z)=S_j(z)+\mathsf{C}(z)^\dagger Y_j(z)\mathsf{C}(z).
\label{eq:lyapunov}
\end{equation}
This is the discrete-time Lyapunov equation, also known as the Stein equation.
Its solution is unique under the stability condition.
Indeed, the difference $\Delta$ between any two solutions satisfies $\Delta=\mathsf{C}(z)^{\dagger n} \Delta\mathsf{C}(z)^n$ for every $n$, which implies $\Delta=0$ as $n\to\infty$.

Vectorization, which stacks the columns of a matrix into a single vector, converts Eq.~\eqref{eq:lyapunov} into a linear system with $d^2$ unknowns, where $d=\dim\mathcal{H}_{\mathcal{C}}$:
\begin{equation}
\operatorname{vec}Y_j(z)
=
\left[I_{d^2}-\mathsf{C}(z)^{\mathsf T}\otimes\mathsf{C}(z)^\dagger\right]^{-1}
\operatorname{vec}S_j(z).
\label{eq:vector-lyapunov}
\end{equation}
The inverse exists under the uniform stability condition.
Thus, the infinite sum can be evaluated by solving a finite linear system.

To identify the dependence on $z$, recall from Eqs.~\eqref{eq:phase-router} and~\eqref{eq:phase-branches} that
\begin{equation}
\mathsf{C}(z)=\mathsf{C}\Gamma(z)=\mathsf{C}_0+z\mathsf{C}_1,    
\end{equation}
where
$\mathsf{C}_0:=\mathsf{C}(I_{\mathcal{C}}-\Pi_s)$ and
$\mathsf{C}_1:=\mathsf{C}\Pi_s$.
Thus, $\mathsf{C}(z)$ is affine in $z$, as are the terminal matrices $\mathsf{T}_j(z)=\mathsf{T}_j\Gamma(z)$.
On the unit circle, complex conjugation replaces $z$ by $z^{-1}$.
Consequently, both $S_j(z)$ and the coefficient matrix in Eq.~\eqref{eq:vector-lyapunov} have Laurent-polynomial entries in $z$.
The inverse coefficient matrix therefore has rational entries, and so does $Y_j(z)$.

The same argument also applies to the coherence kernel $\chi_j(z,w)$, given by Eq.~\eqref{eq:chi}, which describes the transformation of coherences between phase sectors $z$ and $w$.
Define
\begin{equation}
S_j(z,w)=\mathsf{T}_j(w)^\dagger\mathsf{T}_j(z).
\label{eq:mixed-source}
\end{equation}
Then Eq.~\eqref{eq:chi} gives
$\chi_j(z,w)=\bra{\beta}Y_j(z,w)\ket{\beta}$, where
\begin{equation}
Y_j(z,w)=\sum_{n=0}^{\infty}
\bigl(\mathsf{C}(w)^\dagger\bigr)^n
S_j(z,w)\mathsf{C}(z)^n.
\end{equation}
Separating the first term yields the mixed Lyapunov equation
\begin{equation}
Y_j(z,w)
=S_j(z,w)+\mathsf{C}(w)^\dagger Y_j(z,w)\mathsf{C}(z).
\label{eq:mixed-lyapunov}
\end{equation}
For $w=z$, this reduces to Eq.~\eqref{eq:lyapunov}.
Vectorization again converts the equation into a finite linear system, whose coefficient matrix is invertible under uniform stability.
Since its entries and those of the source are polynomials in $z$ and $w^{-1}$ on the unit torus, $\chi_j(z,w)$ is rational in these variables.
Thus, finite controller memory gives rise to rational terminal probabilities and rational coherence kernels.

The controller dimension also bounds the degree of the rational response.
By Cramer's rule, Eq.~\eqref{eq:vector-lyapunov} expresses $f_j(z)$ as a ratio of Laurent polynomials whose exponents lie between $-d^2$ and $d^2$.
If this response depends on $z$ only through $x=(z+z^{-1})/2$, so that $f_j(z)=\widetilde R(x)$, then
\begin{equation}
\operatorname{rdeg}\widetilde R\leq d^2,
\label{eq:memory-degree-main}
\end{equation}
where, for a reduced rational function $\widetilde R=\widetilde{P}/\widetilde{Q}$, we define
$\operatorname{rdeg}\widetilde R:=\max\{\deg \widetilde{P},\deg \widetilde{Q}\}$.
The complete degree argument is given in Appendix~\ref{app:necessity}.

The same Lyapunov structure appears in stable discrete-time linear systems, where it describes the accumulation of quadratic quantities under repeated evolution.
In our protocol, it sums contributions from distinct measurement histories, with the controller memory serving as the finite internal state.
This finite-dimensional structure gives rise to rational terminal probabilities and coherence kernels~\cite{AlaouiMahoney2015,MuscoMusco2017}.

\subsection{Phase inversion symmetry}
\label{subsec:symmetry}

We have shown that finite controller memory produces a rational response in the walk eigenvalue $z$.
To express this response directly in terms of the original Hermitian eigenvalue $x$, we must account for the two-to-one correspondence between $z_\pm(x)$ and $x=\frac{z+z^{-1}}{2}$ for $z\in\{z_\pm(x)\}$.
We require the acceptance probability to be the same for both $z$ and $z^{-1}$:
\begin{equation}
f_{\mathsf{a}}(z)=f_{\mathsf{a}}(z^{-1}),
\qquad z\in\mathbb{T}.
\label{eq:symmetry}
\end{equation}
This phase-inversion symmetry ensures that the walk-level acceptance effect depends only on Hermitian part $X_{\mathcal{W}}$, defined in Eq.~\eqref{eq:X_W}.

A rational function invariant under $z\mapsto z^{-1}$ can be expressed as a rational function of $(z+z^{-1})/2$.
Thus, Eq.~\eqref{eq:symmetry} implies that there is a unique rational function $R$ such that
\begin{equation}
f_{\mathsf{a}}(z)
=R\left(\frac{z+z^{-1}}{2}\right).
\label{eq:descent}
\end{equation}
Because $f_{\mathsf{a}}$ is real-valued on the unit circle, $R$ can be chosen with real coefficients.
Moreover, uniform termination ensures that $f_{\mathsf{a}}$ has no poles on the unit circle.
Consequently, $R$ has no poles on $[-1,1]$ and, as an acceptance probability, satisfies
$0\leq R(x)\leq1$ throughout this interval.
The algebraic argument establishing Eq.~\eqref{eq:descent} is given in Appendix~\ref{app:necessity}.

By Eq.~\eqref{eq:descent}, both walk eigenvalues associated with $x$ give the same acceptance probability:
\[
f_{\mathsf{a}}(z_+(x))=f_{\mathsf{a}}(z_-(x))=R(x).
\]
Substituting this into Eq.~\eqref{eq:system-response}, we obtain
\begin{equation}
\Xi_{\mathsf{a}}(x,x)=R(x).
\end{equation}
The corresponding system-level POVM effect is therefore
\begin{equation}
E_{\mathsf{a}}
=\sum_\lambda R(x_\lambda)P_\lambda
=R(X).
\label{eq:system-rational-effect}
\end{equation}

On the full walk register, the acceptance POVM effect similarly takes the form
\begin{equation} \label{eq:F_a_from_X_W}
    F_{\mathsf{a}}=\sum_\nu R\left(\frac{z_\nu+z_\nu^{-1}}{2}\right)\Omega_\nu
=R(X_{\mathcal{W}}),
\end{equation}
where the last equality follows from $X_{\mathcal{W}}=(W+W^\dagger)/2$ and spectral calculus.
Equation~\eqref{eq:system-rational-effect} also follows from Eq.~\eqref{eq:F_a_from_X_W} using $E_{\mathsf{a}}=\mathcal{E}^\dagger F_{\mathsf{a}}\mathcal{E}$ and the intertwining relation~\eqref{eq:bridge}, which implies $R(X_{\mathcal{W}})\mathcal{E}=\mathcal{E}R(X)$.

The function $R$ is determined by the controller parameters.
We now formalize the requirement that a single controller realize this response for every input operator $X$.
\begin{definition}[Exact finite-memory RUS realization]
\label{def:realization}
A function $R:[-1,1]\to\mathbb{R}$ admits an exact finite-memory RUS realization if there exist a finite-dimensional controller memory $\mathcal{H}_{\mathcal{C}}$, a route projector $\Pi_s$, a normalized initial controller state $\ket{\beta}$, and signal-independent colligation blocks
$\mathsf{C}$, $\mathsf{T}_{\mathsf{a}}$, and $\mathsf{T}_{\mathsf{r}}$
satisfying Eq.~\eqref{eq:first-column-isometry}, such that the continuation dynamics satisfies the uniform termination condition in Eq.~\eqref{eq:termination} and
\begin{equation}
f_{\mathsf{a}}(z)
=R\left(\frac{z+z^{-1}}{2}\right),
\qquad z\in\mathbb{T}.
\label{eq:realization-definition}
\end{equation}
Equivalently, the same controller yields
$F_{\mathsf{a}}=R(X_{\mathcal{W}})$
for every finite-dimensional Hermitian contraction $X$ and its canonical walk $W$.
The controller parameters may depend on $R$, but not on the dimension, spectrum, or eigenvectors of $X$, or on the input state.
\end{definition}

Requiring the response identity on the entire unit circle makes realizability a property of the target function, rather than a fit to a particular finite spectrum.
In the following sections, we characterize the functions that admit such a realization and show how to construct the corresponding controllers.

\subsection{Representation theorem}
\label{subsec:theorem}
We can now formulate the central result of the framework, which characterizes the spectral functions realizable by finite-memory measured RUS dynamics.

\begin{theorem}[RUS representation theorem]
\label{thm:representation}
For a function $R:[-1,1]\to\mathbb{R}$, the following statements are equivalent.

\smallskip
\noindent\textup{(i)}
$R$ admits an exact finite-memory RUS realization in the sense of Definition~\ref{def:realization}.

\smallskip
\noindent\textup{(ii)}
$R$ is a real rational function with no poles on $[-1,1]$ and $0\leq R(x)\leq1$ throughout this interval.
Equivalently, there exist real polynomials $P,Q\in\mathbb{R}[x]$ such that
\begin{equation}
\begin{gathered}
R(x)=\frac{P(x)}{Q(x)},\\
Q(x)>0,\quad 0\leq P(x)\leq Q(x), \quad x\in[-1,1].
\end{gathered}
\label{eq:representation}
\end{equation}

Moreover, if $r=\operatorname{rdeg}(R)\geq1$, there exists a fixed realization requiring at most $r$ controlled-walk queries per complete trial and controller memory dimension at most $3r$.
Conversely, every realization with memory dimension $d=\dim\mathcal{H}_{\mathcal{C}}$ satisfies
\begin{equation}
r\leq d^2.
\label{eq:degree-memory-theorem}
\end{equation}
Consequently, the minimum memory dimension obeys
\begin{equation}
\left\lceil\sqrt{r}\right\rceil
\leq d_{\min}(R)\leq3r,
\qquad r\geq1.
\label{eq:dmin-bound}
\end{equation}
\end{theorem}

We first outline the proof, highlighting its main steps. The technical details are provided in the following subsections.

\paragraph*{Proof of necessity: (i)$\Rightarrow$(ii).}
Uniform termination guarantees that the terminal outcomes form a complete POVM, $F_{\mathsf{a}}+F_{\mathsf{r}}=I_{\mathcal{W}}$.
Consequently, the acceptance response satisfies $0\leq f_{\mathsf{a}}(z)\leq1$ for every $z\in\mathbb{T}$.

The vectorized Lyapunov equation \eqref{eq:vector-lyapunov} expresses $f_{\mathsf{a}}(z)$ through a finite-dimensional matrix resolvent. Since the continuation operator is affine in $z$, its coefficient matrix and source are Laurent-polynomial on the unit circle.
Thus, $f_{\mathsf{a}}(z)$ is rational in $z$, and uniform stability excludes poles on $\mathbb{T}$.

By Definition~\ref{def:realization}, the response is invariant under $z\mapsto z^{-1}$.
The fixed-field argument of Sec.~\ref{subsec:symmetry} therefore implies that $R$ is a real rational function of $x=(z+z^{-1})/2$. 
Since this map covers $[-1,1]$, the probability bounds and absence of poles descend to the entire Hermitian signal interval.

Writing $R=P/Q$ in reduced form, its denominator has no zeros on $[-1,1]$ and therefore has a constant sign. 
Choosing $Q>0$ gives $0\leq P\leq Q$, establishing Eq.~\eqref{eq:representation}.

Finally, the Laurent-degree bound derived in Sec.~\ref{subsec:lyapunov} gives $r\leq d^2$.
The complete degree argument is provided in Appendix~\ref{app:necessity}.

\paragraph*{Constructive proof of (ii)$\Rightarrow$(i).}
The converse follows by explicitly compiling any function satisfying Eq.~\eqref{eq:representation} into a measured RUS protocol.

The key idea is to represent the target rational function as the eventual acceptance probability of a repeated trial. 
Let $Q_{\max}=\max_{x\in[-1,1]}Q(x)$ and define the polynomial trial probabilities
\begin{equation}
\begin{aligned}
p_{\mathsf{a}}(x)=\frac{P(x)}{Q_{\max}}, &\quad 
p_{\mathsf{r}}(x)=\frac{Q(x)-P(x)}{Q_{\max}},\\
p_{\mathsf{c}}(x)&=1-\frac{Q(x)}{Q_{\max}}.
\end{aligned}
\label{eq:representation-trial}
\end{equation}
They are nonnegative and sum to unity.
Furthermore, $Q(x)>0$ on the compact interval, so the probability of terminating a complete trial is uniformly bounded away from zero.

If continuation repeats the same trial while preserving the spectral sector, the eventual acceptance probability becomes
\begin{equation}
\begin{aligned}
\sum_{n=0}^{\infty}
p_{\mathsf{c}}(x)^n p_{\mathsf{a}}(x)
=\frac{p_{\mathsf{a}}(x)}
{1-p_{\mathsf{c}}(x)}=\frac{P(x)}{Q(x)}
=R(x).
\end{aligned}
\label{eq:representation-renewal}
\end{equation}
The rational denominator therefore arises directly from the repeated measured dynamics. Sec.~\ref{sec:compiler} shows how to implement these trial probabilities with controlled-$W$ queries.

The construction realizes the required response uniformly over the unit circle and guarantees termination. 
Since the trial polynomials have degree at most $r$, the realization requires at most $r$ controlled-walk queries per complete trial and memory dimension $d\leq3r$, as established in Sec.~\ref{subsec:cost}.
Together with the necessity bound, this proves Eq.~\eqref{eq:dmin-bound} and completes the proof. \hfill$\square$

We conclude with a special case and the operational interpretation of the theorem.
For constant functions ($r=0$), a signal-independent terminal measurement suffices, requiring no walk queries.
To interpret this characterization operationally, recall that a realization of $R$ has acceptance effects
\begin{equation}
F_{\mathsf{a}}=R(X_{\mathcal{W}}), \qquad
E_{\mathsf{a}}=\mathcal{E}^\dagger F_{\mathsf{a}}\mathcal{E}=R(X).
\label{eq:representation-effects}
\end{equation}
Thus, for a system input $\rho_{\mathcal{S}}$, the acceptance probability is
\begin{equation}
\Pr(\mathsf{a}\mid\rho_{\mathcal{S}})
=\operatorname{Tr}\left[R(X)\rho_{\mathcal{S}}\right].
\label{eq:representation-acceptance}
\end{equation}
The theorem characterizes exactly which functions $R$ admit such a realization.

\subsection{Termination-time and expected query cost}
\label{subsec:cost}

The representation theorem guarantees a finite-memory realization of every admissible rational response.
We now quantify the resources of the constructive protocol, distinguishing the cost of a complete trial from the number of trials required for termination.

As outlined in the proof of Theorem~\ref{thm:representation}, the compiler decomposes each trial into a finite sequence of elementary operations $U_1,\ldots,U_L$.
To implement this sequence using one fixed router, one fixed colligation, and the same flag measurement after every round, the section index is stored coherently in a finite register, which we call the clock register.
The first $L-1$ rounds return the continuation outcome deterministically and advance the clock.
The final round produces acceptance, rejection, or continuation, with the latter resetting the clock for another trial.
The explicit clocked isometry and its unitary completion are given in Appendix~\ref{app:clock}.
This construction requires at most $3L$ memory dimensions and $L$ controlled-walk queries per complete trial.

Consider a target function $R(x)=P(x)/Q(x)$ satisfying
Eq.~\eqref{eq:representation}.
The construction permits a real polynomial multiplier $S(x)$ satisfying
\begin{equation}
0<S(x)Q(x)\leq1, \quad x\in[-1,1].
\label{eq:multiplier-condition}
\end{equation}
The corresponding one-trial probabilities are
\begin{equation}
\begin{aligned}
p_{\mathsf a}(x)&=S(x)P(x),\\
p_{\mathsf r}(x)&=S(x)[Q(x)-P(x)],\\
p_{\mathsf c}(x)&=1-S(x)Q(x).
\end{aligned}
\label{eq:cost-trial-probabilities}
\end{equation}
Their sum is unity, while the probability of terminating a complete trial is
\begin{equation}
\tau(x)=p_{\mathsf a}(x)+p_{\mathsf r}(x)
=S(x)Q(x).
\label{eq:termination-probability}
\end{equation}
The strict positivity of $\tau(x)$ on the compact signal interval guarantees
\begin{equation}
\tau_{\min} = \min_{x\in[-1,1]}\tau(x)>0.
\label{eq:tau-min}
\end{equation}

Let $N$ denote the number of complete trials up to and including termination, and let $\mu_{\rho_{\mathcal S}}$ be the spectral measure of the input state with respect to $X$, given by
\begin{equation}
\mu_{\rho_{\mathcal S}}
=\sum_\lambda
\operatorname{Tr}(P_\lambda\rho_{\mathcal S})\,\delta_{x_\lambda},
\end{equation}
where $\delta_{x_\lambda}$ is the unit point mass at $x_\lambda$.
The survival probability after $n$ trials is
\begin{equation}
\begin{aligned}
\Pr(N>n|\rho_{\mathcal S})
&=
\int_{-1}^{1}
[1-\tau(x)]^n\,
d\mu_{\rho_{\mathcal S}}(x)\\
&\leq(1-\tau_{\min})^n.
\end{aligned}
\label{eq:tail}
\end{equation}
Thus, the compiled recurrence has an exponentially decaying termination tail.
Its expected number of complete trials is
\begin{equation}
\mathbb E[N|\rho_{\mathcal S}]
=
\int_{-1}^{1}
\frac{d\mu_{\rho_{\mathcal S}}(x)}
{S(x)Q(x)}
\leq\frac{1}{\tau_{\min}}.
\label{eq:expected-trials}
\end{equation}

For a clocked trial of length $L(S)$, the number of elementary rounds is $T=L(S)N$.
Since each scheduled section uses at most one controlled-walk query, the expected query cost of a terminal run is bounded by
\begin{equation}
C_{\mathrm{run}}(S;\rho_{\mathcal S}) = L(S)\int_{-1}^{1} \frac{d\mu_{\rho_{\mathcal S}}(x)} {S(x)Q(x)} \leq \frac{L(S)}{\tau_{\min}},
\label{eq:run-cost}
\end{equation}
where the first expression counts the scheduled query slots and
\begin{equation}
L(S)= \max\left\{ \deg(SP),\, \deg[S(Q-P)],\, \deg(1-SQ) \right\},
\label{eq:trial-degree}
\end{equation}
with identically zero branches omitted.

A particularly simple choice is $S(x)=1/Q_{\max}$, where $Q_{\max}=\max_{x\in[-1,1]}Q(x)$.
It gives $L\leq r=\operatorname{rdeg}(R)$, controller dimension $d\leq3r$, and
\begin{equation}
C_{\mathrm{run}}(\rho_{\mathcal S}) = LQ_{\max} \int_{-1}^{1} \frac{d\mu_{\rho_{\mathcal S}}(x)}{Q(x)}.
\label{eq:constant-multiplier-cost}
\end{equation}
The query cost therefore depends on both the polynomial trial length and the probability that a trial terminates for the input spectrum.

The eventual acceptance probability is independent of the multiplier $S$:
\begin{equation}
\alpha_R(\rho_{\mathcal S})= \operatorname{Tr} \left[R(X)\rho_{\mathcal S}\right].
\label{eq:cost-acceptance}
\end{equation}
If rejection is followed by a fresh preparation of the original input state and an independent run, the expected query cost per accepted outcome becomes
\begin{equation}
C_{\mathsf a}(S;\rho_{\mathcal S})= \frac{C_{\mathrm{run}}(S;\rho_{\mathcal S})} {\alpha_R(\rho_{\mathcal S})},
\label{eq:accepted-cost}
\end{equation}
provided $\alpha_R(\rho_{\mathcal S})>0$.
The cost of input re-preparation is not included in this expression.

The choice of $S$ changes the trial degree and termination profile, while leaving the eventual acceptance effect $R(X)$ unchanged.
Minimizing the expected query cost amounts to balancing the complexity $L(S)$ against the spectral average of $1/[S(x)Q(x)]$.
Appendix~\ref{app:optimization} develops a bounded-degree optimization formulation using exact polynomial positivity certificates.

\section{Constructive compilation}
\label{sec:compiler}

We now complete the constructive argument of Theorem~\ref{thm:representation} by building a quantum circuit for the polynomial trial probabilities introduced above.
The construction proceeds in three steps.
First, Fej\'er--Riesz factorization produces polynomial amplitudes whose squared magnitudes reproduce the desired probabilities.
We then realize these amplitudes using only forward controlled-walk queries and signal-independent unitaries.
Finally, a clock register embeds the resulting sequence into a fixed controller that repeats the trial whenever the continuation outcome occurs.

\subsection{Phase lift and spectral factorization}
\label{sec:spectral-factor}
Consider the polynomial trial probabilities
$p_j(x)$,
$j\in\{\mathsf a,\mathsf r,\mathsf c\}$,
defined in Eq.~\eqref{eq:representation-trial}.
To implement them using the walk operator, we lift them to the unit circle:
\begin{equation}
\widehat p_j(z) = p_j\left(\frac{z+z^{-1}}{2}\right).
\label{eq:phase-lift}
\end{equation}
Each $\widehat p_j$ is a nonnegative trigonometric polynomial on $\mathbb T$.
If $p_j$ has degree $L_j$, its lift has Laurent support in $[-L_j,L_j]$.
Their finite polynomial degree allows us to implement the amplitudes with finitely many walk queries.

By the Fej\'er--Riesz theorem, there exist causal polynomials $g_j(z)$ such that
\begin{equation}
\widehat p_j(z)=|g_j(z)|^2, \quad \sum_j|g_j(z)|^2=1, \quad z\in\mathbb T.
\label{eq:factors}
\end{equation}
The degree of each factor is at most $L_j$, and an identically zero branch may be represented by $g_j=0$.
This step is analogous to the conversion of a nonnegative power response into a causal finite impulse response factor in classical signal-processing theory~\cite{EfrosEtAl2007,Rall2020,Roggero2020}.

Introduce a three-dimensional branch space with orthonormal basis $\{\ket{\mathsf a},\ket{\mathsf r}, \ket{\mathsf c}\}$ and define the polynomial column
\begin{equation}
\ket{g(z)}=g_{\mathsf a}(z)\ket{\mathsf a}+g_{\mathsf r}(z)\ket{\mathsf r}+g_{\mathsf c}(z)\ket{\mathsf c}.
\label{eq:inner-column}
\end{equation}
Equation~\eqref{eq:factors} implies $\langle g(z)|g(z)\rangle=1$, for $z\in\mathbb T$.
Thus, $\ket{g(z)}$ is an inner, or lossless, polynomial column.
Because the squared amplitudes sum to one, they can be combined into a sequence of unitary operations~\cite{Kailath1980,ZhouDoyleGlover1996}.
Since the phase lift preserves the polynomial degree and the Fej\'er--Riesz factors can be chosen with minimum degree, we have $\deg g_j=\deg p_j$ for every nonzero branch. 
Consequently, the trial length $L=L(S)$, already defined in Eq. (70), is also the maximum degree of the polynomial amplitudes. 
The resulting inner column can therefore be synthesized using exactly $L$ elementary degree-stripping sections.
The task is now to construct this column using at most $L$ forward controlled-walk queries.

\subsection{Degree-stripping realization}
\label{sec:degree-strip-main}

We construct the circuit by removing the highest power of $z$ from the polynomial vector, one step at a time.
The resulting factorization is
\begin{equation}
\ket{g(z)}= E_{\Lambda_L}(z)\cdots E_{\Lambda_1}(z)\ket{u},
\label{eq:factorization}
\end{equation}
where $\ket{u}\in\mathbb C^3$ is a constant unit vector and
\begin{equation}
E_\Lambda(z) = I_3-\Lambda+z\Lambda
\label{eq:elementary-factor}
\end{equation}
for rank-one orthogonal projectors $\Lambda$.
Each factor is unitary on the unit circle and therefore preserves the inner condition.
The construction is closely related to lossless lattice factorizations and Schur-type realizations~\cite{DritschelRovnyak2010,Dumitrescu2007}.

To derive the recursion, write the current inner column of degree $\ell$ as
\begin{equation}
\ket{g^{(\ell)}(z)} = \sum_{k=0}^{\ell} z^k\ket{g_k^{(\ell)}}.
\label{eq:inner-current-main}
\end{equation}
The leading coefficient determines the rank-one projector
\begin{equation}
\Lambda_\ell = \frac{\ket{g_\ell^{(\ell)}}\bra{g_\ell^{(\ell)}}}{\langle g_\ell^{(\ell)}|g_\ell^{(\ell)}\rangle}.
\label{eq:stripping-projector-main}
\end{equation}
The reduced column is
\begin{equation}
\ket{g^{(\ell-1)}(z)}=\left(I_3-\Lambda_\ell+z^{-1}\Lambda_\ell \right)\ket{g^{(\ell)}(z)}.
\label{eq:stripping-step-main}
\end{equation}
Although Eq.~(\ref{eq:stripping-step-main}) contains $z^{-1}$, the term with a negative power vanishes because the polynomial column is normalized.
Indeed, the highest Fourier coefficient of $\langle g^{(\ell)}(z)|g^{(\ell)}(z)\rangle=1$ gives $\langle g_\ell^{(\ell)} |g_0^{(\ell)}\rangle=0$.
Therefore, $\Lambda_\ell\ket{g_0^{(\ell)}}=0$, so no negative power of $z$ appears.
At the same time, $(I_3-\Lambda_\ell) \ket{g_\ell^{(\ell)}}=0$, which removes the highest positive power.
The degree decreases by one while normalization is preserved.

Iteration eventually produces the constant vector $\ket{u}$.
Reversing the steps yields Eq.~\eqref{eq:factorization}.
The complete coefficient recursion and proof are given in Appendix~\ref{app:compiler}.

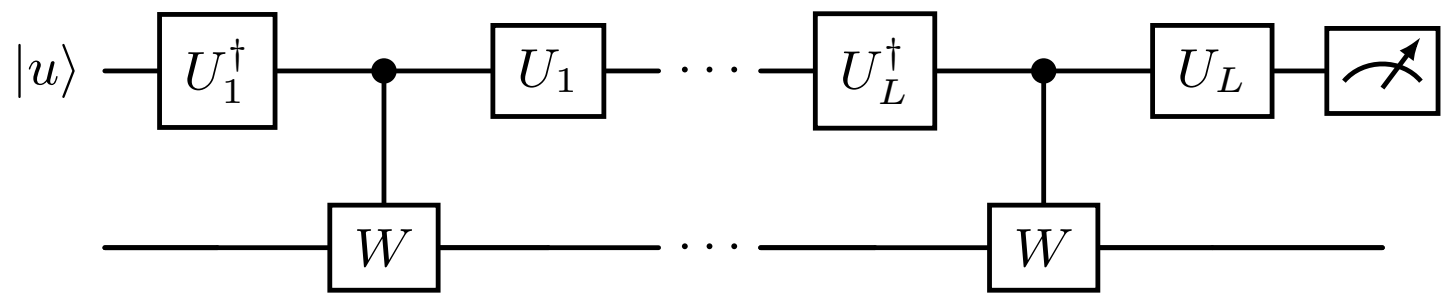
\begin{figure*}[t]
\centering
\begin{minipage}[t]{0.54\textwidth}
\centering
\resizebox{\linewidth}{!}{%
\begin{quantikz}[
row sep=0.45cm,
column sep=0.32cm
]
\lstick{$\ket{u}$}
& \gate{U_1^\dagger}
& \ctrl{1}
& \gate{U_1}
& ~\cdots~
& \gate{U_L^\dagger}
& \ctrl{1}
& \gate{U_L}
& \meter{} \\
\lstick{}
& \qw
& \gate{W}
& \qw
& ~\cdots~
& \qw
& \gate{W}
& \qw
& \qw
\end{quantikz}}
\end{minipage}
\caption{Compilation of one polynomial trial. 
Fej\'er--Riesz factorization and degree stripping produce a finite sequence of controlled-walk sections with branch amplitudes $g_{\mathsf a}(W)$, $g_{\mathsf r}(W)$, and $g_{\mathsf c}(W)$.
A finite clock embeds this sequence into the fixed-round measured RUS architecture.}
\label{fig:circuits}
\end{figure*}

Each elementary factor has a direct quantum implementation.
Choose a fixed route state $\ket{s}$ and a unitary $U_\ell$ satisfying
\begin{equation}
\Lambda_\ell= U_\ell\ket{s}\bra{s}U_\ell^\dagger.
\label{eq:route-projector-factor}
\end{equation}
The corresponding walk-dependent section is
\begin{equation}
E_{\Lambda_\ell}(W)= (U_\ell\otimes I_{\mathcal W}) \Gamma_{s,W} (U_\ell^\dagger\otimes I_{\mathcal W}),
\label{eq:section}
\end{equation}
where
\begin{equation}
\Gamma_{s,W} = (I_3-\ket{s}\bra{s})\otimes I_{\mathcal W} + \ket{s}\bra{s}\otimes W.
\label{eq:trial-router}
\end{equation}
Thus, each section requires exactly one forward controlled-$W$ query, together with signal-independent unitaries acting on the branch register (Fig.~\ref{fig:circuits}).
Although the algebraic stripping recursion contains $z^{-1}$, the reconstructed circuit uses only the forward factors $E_{\Lambda_\ell}(W)$.

After the $L$ sections, measurement of the branch register implements the three walk operators
\begin{equation}
G_j=g_j(W), \quad G_j^\dagger G_j=p_j(X_{\mathcal W}).
\label{eq:compiled-trial-operators}
\end{equation}
Their completeness follows directly from the normalization of the polynomial column:
\begin{equation}
\sum_jG_j^\dagger G_j=I_{\mathcal W}.
\label{eq:compiled-trial-completeness}
\end{equation}

If the continuation branch is obtained, the retained walk register enters another identical trial.
The first acceptance after $n$ continuing trials therefore has Kraus operator $G_{\mathsf a}G_{\mathsf c}^n$.
All branch operators are polynomials in the same unitary $W$ and hence commute. 
Their cumulative accepting effect is
\begin{equation}
\begin{aligned}
F_{\mathsf a}
=
\sum_{n=0}^{\infty}
G_{\mathsf c}^{\dagger n}
G_{\mathsf a}^\dagger G_{\mathsf a}
G_{\mathsf c}^{n}
&=
p_{\mathsf a}(X_{\mathcal W})
\left[
I_{\mathcal W}
-p_{\mathsf c}(X_{\mathcal W})
\right]^{-1} \\
&=R(X_{\mathcal W}).
\end{aligned}
\label{eq:compiled-effect}
\end{equation}
This verifies the constructive representation at the level of the complete polynomial trial.

Finally, the scheduled sequence of elementary sections is promoted to the fixed-round RUS architecture.
A finite clock stores the section number coherently and advances after each elementary round.
The measured flag returns continuation deterministically during the first $L-1$ rounds. 
At the final section, it produces acceptance, rejection, or continuation, with the latter resetting the clock to its initial state without discarding the walk register.

The resulting protocol uses one fixed router, one fixed unitary colligation, and the same flag measurement after every elementary round. 
The explicit clocked isometry and its unitary completion are given in Appendix~\ref{app:clock}; the associated memory, termination, and query bounds are established in Sec.~\ref{subsec:cost}.

The compilation thus connects rational spectral filtering to classical spectral factorization and lossless realization theory.
Fej\'er--Riesz factorization produces polynomial amplitudes that reproduce the desired trial probabilities.
We then construct a finite circuit for these amplitudes using controlled-walk queries and signal-independent unitaries. 
Finally, a clock register allows a fixed controller to execute this circuit and repeat the trial whenever continuation is measured.
Importantly, spectral factorization also determines the phases of the branch amplitudes, and hence the coherence structure of the accepted quantum operation.
Different compilations of the same rational effect may therefore produce different coherence kernels, as discussed in Sec.~\ref{subsec:kernel-design}.

\subsection{Kernel factorization and channel design}
\label{subsec:kernel-design}

The trial amplitudes determine not only the acceptance probability but also the output quantum state. 
We now characterize that state transformation.

For the compiled renewal protocol, the first acceptance after $n$ continuing trials applies the walk operator $G_{\mathsf a}G_{\mathsf c}^{n}$, where $G_j=g_j(W)$.
Consequently, the phase-level coherence kernel introduced in Eq.~\eqref{eq:chi} takes the form
\begin{equation}
\begin{aligned}
\chi_{\mathsf a}(z,w)
&=
\sum_{n=0}^{\infty}
g_{\mathsf a}(z)
\overline{g_{\mathsf a}(w)}
\left[
g_{\mathsf c}(z)
\overline{g_{\mathsf c}(w)}
\right]^n\\
&=
\frac{
g_{\mathsf a}(z)
\overline{g_{\mathsf a}(w)}
}{
1-g_{\mathsf c}(z)
\overline{g_{\mathsf c}(w)}
}.
\end{aligned}
\label{eq:chi-renewal}
\end{equation}
Uniform termination guarantees convergence of this geometric series.
Its diagonal reproduces the rational response,
\begin{equation}
\chi_{\mathsf a}(z,z) = \frac{|g_{\mathsf a}(z)|^2} {1-|g_{\mathsf c}(z)|^2} = R\left(\frac{z+z^{-1}}{2}\right).
\label{eq:renewal-kernel-diagonal}
\end{equation}
Thus, while the rational effect is fixed by the trial probabilities, the off-diagonal kernel retains information about the phases and spectral variation of the compiled amplitudes.

To obtain the accepted operation on the original system, recall the canonical phase correspondence $z_\pm(x)=e^{\pm i\arccos x}$.
After embedding the system input and tracing out the block-encoding qubit, the system kernel is
\begin{equation}
\Xi_{\mathsf a}(x,y) = \frac12\sum_{\sigma\in\{+,-\}} \chi_{\mathsf a} \bigl(z_\sigma(x),z_\sigma(y)\bigr).
\label{eq:compiled-system-kernel}
\end{equation}
In particular, $\Xi_{\mathsf a}(x,x)=R(x)$, consistently with the acceptance effect established in Eq.~\eqref{eq:compiled-effect}.

The structure of this kernel becomes transparent through a spectral Kraus representation.
Let $n$ label the number of continuing trials before acceptance and $\sigma$ the canonical walk-phase branch. 
Define
\begin{equation}
\begin{aligned}
L_{n,\sigma}
&= \sum_\lambda \ell_{n,\sigma}(x_\lambda)P_\lambda,\\ 
\ell_{n,\sigma}(x)
&=\frac{g_{\mathsf a}(z_\sigma(x)) g_{\mathsf c}(z_\sigma(x))^n}{\sqrt2}.
\end{aligned}
\label{eq:spectral-kraus}
\end{equation}
The accepted system operation is then
\begin{equation}
\Psi_{\mathsf a}(\rho_{\mathcal S}) = \sum_{n=0}^{\infty} \sum_{\sigma\in\{+,-\}} L_{n,\sigma}\rho_{\mathcal S} L_{n,\sigma}^\dagger.
\label{eq:compiled-system-operation}
\end{equation}
The derivation from the canonical walk and the measured history decomposition is given in Appendix~\ref{app:channel}.

Collect these amplitudes into the spectral
history vector
\begin{equation}
\ket{\zeta(x)} = \sum_{n=0}^{\infty} \sum_{\sigma\in\{+,-\}} \ell_{n,\sigma}(x)\ket{n,\sigma}.
\label{eq:spectral-feature-vector}
\end{equation}
The system coherence kernel is precisely their Gram kernel:
\begin{equation}
\Xi_{\mathsf a}(x,y) = \langle\zeta(y)|\zeta(x)\rangle, \quad \|\zeta(x)\|^2=R(x).
\label{eq:kraus-kernel}
\end{equation}
The acceptance effect therefore fixes the norms of the history vectors, while their mutual overlaps determine the coherences retained between different spectral sectors.
In particular, complete positivity is encoded by the positive semidefiniteness of $\Xi_{\mathsf a}$ on every finite spectral set.

This Gram representation motivates a separation between spectral weighting and coherence transformation. 
On the acceptance support
\begin{equation}
\Sigma_R = \{x\in\operatorname{spec}(X):R(x)>0\},
\label{eq:kernel-support}
\end{equation}
define the normalized correlation kernel
\begin{equation}
c_R(x,y) = \frac{\Xi_{\mathsf a}(x,y)} {\sqrt{R(x)R(y)}}.
\label{eq:correlation-kernel}
\end{equation}
Writing $\ket{\zeta(x)}=\sqrt{R(x)}\ket{u(x)}$ with $\|u(x)\|=1$, we obtain
\begin{equation}
c_R(x,y)=\langle u(y)|u(x)\rangle.
\label{eq:normalized-gram}
\end{equation}
Hence $c_R$ is a positive-semidefinite kernel with unit diagonal.
If $R(x)=0$, positivity implies $\Xi_{\mathsf a}(x,y)=0$ for every $y$, so the normalization is needed only on $\Sigma_R$.

Define the spectral Schur channel by
\begin{equation}
\mathcal D_{c_R} (P_\lambda A P_\kappa) = c_R(x_\lambda,x_\kappa) P_\lambda A P_\kappa
\label{eq:schur-channel}
\end{equation}
on the support of $R(X)$.
Positivity and unit diagonal of $c_R$ ensure that this map is completely positive and trace preserving on that subspace.

\begin{theorem}[Channel factorization]
\label{thm:channel-factorization}
Every spectrally aligned accepted operation with acceptance effect $R(X)$ admits the factorization
\begin{equation}
\Psi_{\mathsf a}(\rho_{\mathcal S}) = \mathcal D_{c_R} \left(R(X)^{1/2}\rho_{\mathcal S}R(X)^{1/2} \right).
\label{eq:channel-factorization}
\end{equation}
Conversely, every positive-semidefinite, unit-diagonal kernel $c$ on $\Sigma_R$ defines through this factorization a completely positive, trace-nonincreasing spectral operation with effect $R(X)$.
\end{theorem}

\noindent
\emph{Proof.}
The $(\lambda,\kappa)$ spectral block first acquires the factor $\sqrt{R(x_\lambda)R(x_\kappa)}$ and is then multiplied by $c_R(x_\lambda,x_\kappa)$.
Their product reproduces $\Xi_{\mathsf a}(x_\lambda,x_\kappa)$, establishing the factorization.
Conversely, a positive-semidefinite correlation kernel defines a completely positive Schur channel, and its unit diagonal guarantees that the resulting operation has effect $R(X)$.
\hfill$\square$

The theorem separates two complementary aspects of spectral filtering. The rational effect $R(X)$ determines terminal probabilities and accepted spectral populations, whereas $c_R$ determines the coherence retained by the accepted operation.
Importantly, the converse characterizes valid spectral quantum operations;
realizability by a particular finite-memory RUS architecture imposes additional compilation constraints.

If all nonzero history vectors are collinear, the accepted operation has one spectral Kraus operator.

\begin{theorem}[Rank-one kernel]
\label{thm:rank-one-kernel}
On a nonempty acceptance support $\Sigma\subseteq\Sigma_R$, the restricted kernel $\Xi_{\mathsf a}$ has rank one iff there exists a scalar amplitude $k(x)$ such that
\begin{equation}
\Xi_{\mathsf a}(x,y) = k(x)\overline{k(y)}, \quad |k(x)|^2=R(x).
\label{eq:rank-one-kernel}
\end{equation}
Equivalently, the accepted operation has a single spectral Kraus operator,
\begin{equation}
\Psi_{\mathsf a}(\rho_{\mathcal S}) = k(X)\rho_{\mathcal S}k(X)^\dagger
\label{eq:coherent-filter-channel}
\end{equation}
on that support.
\end{theorem}

\noindent
\emph{Proof.}
The Gram kernel has rank one exactly when its nonzero feature vectors are collinear.
Writing $\ket{\zeta(x)}=k(x)\ket{v}$ for a fixed unit vector $\ket{v}$ gives Eq.~\eqref{eq:rank-one-kernel}.
Substitution into the spectral channel representation proves the single-Kraus form. 
The converse is immediate. \hfill$\square$

This rank-one limit corresponds to coherent spectral filtering: no distinguishable which-spectrum information is retained in the unresolved environment record.
A successful QSP/QSVT postselection with spectral amplitude $p(X)$ provides an example with $k=p$ and $R(x)=|p(x)|^2$.
For a finite spectrum, the rank of the matrix $[\Xi_{\mathsf a}(x_\lambda,x_\kappa)]$ also equals the minimum number of spectral Kraus features needed to represent the accepted operation.

We can now return to the compiled renewal amplitudes and quantify their effect on spectral coherence. 
Wherever the phase-level acceptance probability is nonzero, define
\begin{equation}
c_{\mathrm{ph}}(z,w) = \frac{\chi_{\mathsf a}(z,w)}{\sqrt{f_{\mathsf a}(z)f_{\mathsf a}(w)}}.
\label{eq:phase-correlation-kernel}
\end{equation}
Using Eq.~\eqref{eq:chi-renewal}, this becomes
\begin{equation}
\begin{aligned}
c_{\mathrm{ph}}(z,w)
={}&
e^{i[\varphi_{\mathsf a}(z)
-\varphi_{\mathsf a}(w)]}\\
&\times
\frac{
\sqrt{
[1-|g_{\mathsf c}(z)|^2]
[1-|g_{\mathsf c}(w)|^2]
}
}{
1-g_{\mathsf c}(z)
\overline{g_{\mathsf c}(w)}
},
\end{aligned}
\label{eq:renewal-correlation}
\end{equation}
where $\varphi_{\mathsf a}(z)$ denotes the phase of $g_{\mathsf a}(z)$.
A direct calculation yields
\begin{equation}
1-|c_{\mathrm{ph}}(z,w)|^2= \frac{ |g_{\mathsf c}(z)-g_{\mathsf c}(w)|^2 }{ |1-g_{\mathsf c}(z) \overline{g_{\mathsf c}(w)}|^2}.
\label{eq:coherence-loss}
\end{equation}
Eq.~(\ref{eq:coherence-loss}) shows how the stopping record affects coherence: different continuation amplitudes make the corresponding histories distinguishable and reduce their overlap

The system-level kernel additionally combines the two canonical walk phases, as in Eq.~\eqref{eq:compiled-system-kernel}.
Consequently, perfect phase-level coherence does not by itself guarantee a rank-one system kernel: the discarded block-encoding qubit may retain further spectral information.

These observations suggest a channel-aware approach to rational filter compilation.
Besides minimizing polynomial degree and expected query cost, one may optimize the phases and spectral dependence of the trial amplitudes to control the correlation kernel $c_R(x,y)$.
In particular, preserving coherence within a target spectral band amounts to keeping the corresponding history vectors nearly collinear.

For applications based on terminal probabilities or spectral sampling, the acceptance effect $R(X)$ is sufficient. 
When the accepted state itself is the computational output, the full kernel determines the transformation of spectral superpositions. 
Thus, the same compilation framework naturally extends from rational effect synthesis to the design of spectral quantum operations.
The robustness of the resulting rational spectral transformation against perturbations of the walk oracle and the underlying Hermitian signal is analyzed in Appendix~\ref{app:robustness}.

\section{Applications}
\label{sec:applications}

\begin{figure*}[t]
\centering
\includegraphics[width=0.8\textwidth]{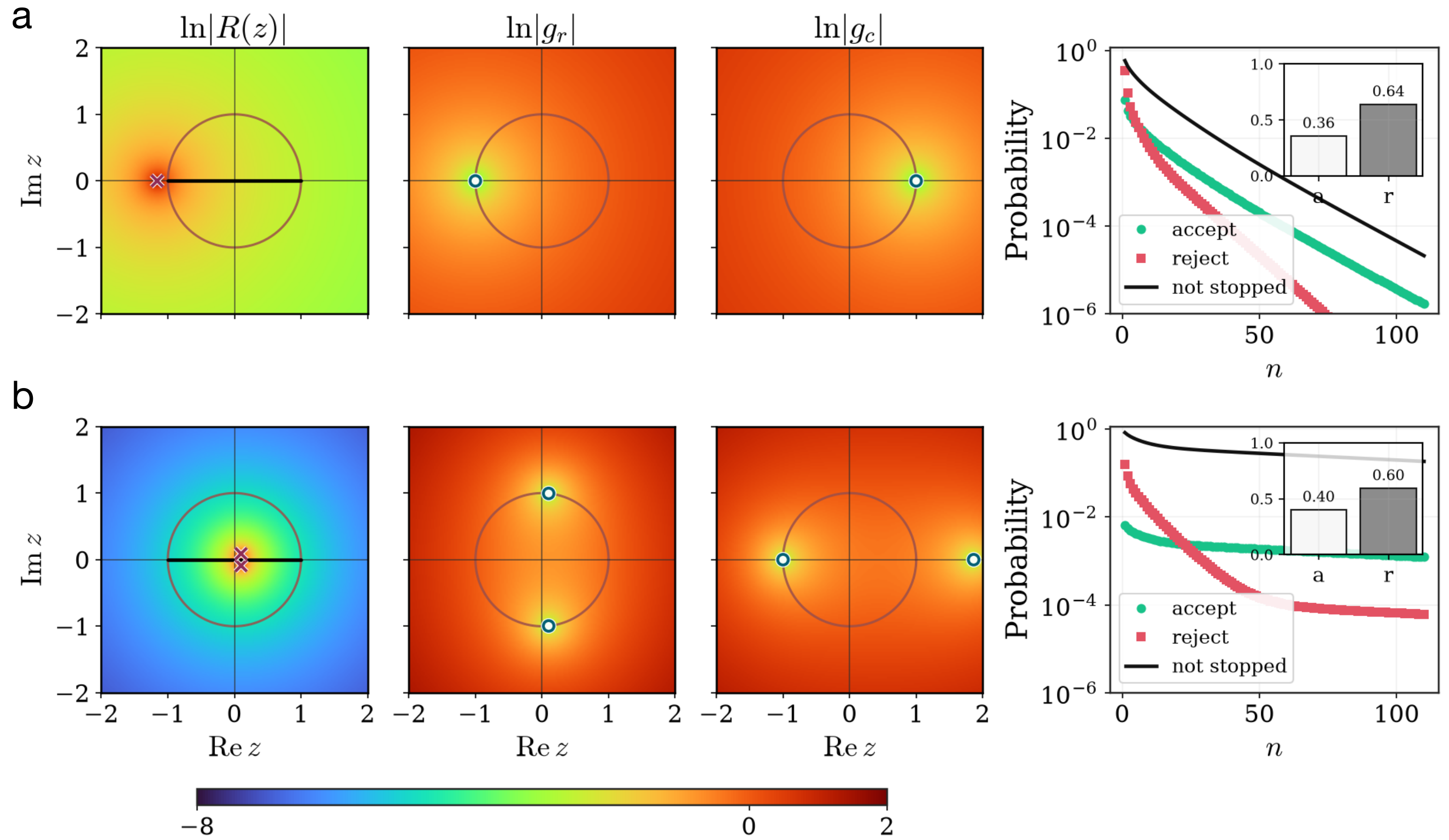}
\vspace{-6pt}
\caption{
Analytic structure and first-hitting statistics of the two application filters: (a) regularized resolvent and (b) Lorentzian spectral window.
The first three columns display $\ln|R(x(z))|$, $\ln|g_{\mathsf r}(z)|$, and $\ln|g_{\mathsf c}(z)|$ in the complex walk-phase plane.
The right column shows the first-hitting acceptance and rejection probabilities together with the nonterminated probability.
The insets display the cumulative terminal acceptance and rejection weights.
}
\label{fig:application-filters}
\end{figure*}

We illustrate the framework with two applications: regularized resolvent filtering and Lorentzian spectroscopy.
The low rational degree of these filters allows us to construct explicit RUS circuits and evaluate their expected query costs.
We focus on sampling from the reweighted spectral distribution and, for input states commuting with the signal operator, preparing the corresponding filtered quantum ensembles.
In both applications, we count the expected number of controlled-walk queries per accepted output, including independent restarts from the original input after rejection.

\subsection{Comparison model}
\label{sec:comparison-model}
We consider a common signal oracle $W$ and assume access to fresh copies of the same input state $\rho_{\mathcal S}$.
Inverse walk calls, where used by the polynomial baseline, are assigned the unit query cost.
The QSP/QSVT baseline consists of a fixed-depth polynomial spectral transformation followed by terminal postselection.
A failed postselection requires a fresh input copy.

We compare against implementations whose acceptance response $\widetilde R_d(x)$ approximates the target response $R(x)$ uniformly,
\begin{equation}
\sup_{x\in[-1,1]}
|\widetilde R_d(x)-R(x)|
\leq\tau,
\label{eq:comparison-accuracy}
\end{equation}
where $d$ denotes the number of signal queries in the fixed-depth implementation.
The approximation is required to be a valid acceptance response, with values in $[0,1]$.

The polynomial degree of an acceptance response need not equal the
number $d$ of signal queries. 
If a $d$-query QSVT circuit implements an amplitude $q_d(x)$, its acceptance response is $\widetilde R_d(x)=|q_d(x)|^2$, a polynomial of degree at most $2d$.
The same degree bound holds for a phase-symmetric acceptance response of a $d$-query walk-based QSVT circuit: its amplitudes are Laurent polynomials in $z$ supported in $[-d,d]$, and their probabilities have Laurent degree at most $2d$.
Conversely, every polynomial $p_D$ of degree $D$ with $0\leq p_D(x)\leq1$ admits a fixed-depth acceptance realization with at most $D$ controlled-walk queries. 
Fej\'er--Riesz factorization of the phase lifts of $p_D$ and $1-p_D$, followed by the degree-stripping construction of Sec.~\ref{sec:compiler}, produces the required two-outcome circuit.
If an unrestricted approximant $p_D$ has error at most $\varepsilon$
against $0\leq R\leq1$, replacing it with $(p_D+\varepsilon)/(1+2\varepsilon)$ makes it a valid response, with error at most $2\varepsilon$ and unchanged degree.
Thus, the minimum polynomial approximation degree and the required fixed-depth query count agree up to constant factors in the regimes considered below.

The end-to-end resource is
\begin{equation}
C_{\mathrm{out}} :=\mathbb E[N_W],
\label{eq:comparison-total}
\end{equation}
where $N_W$ counts all controlled-walk queries, including those spent on continuation trials and rejected runs, until the first accepted output is obtained.
For the RUS protocol, the restart identity derived in Sec.~\ref{subsec:cost} gives
\begin{equation}
C_{\mathrm{out}}^{\mathrm{RUS}} = \frac{C_{\mathrm{run}}} {\alpha_R(\rho_{\mathcal S})}, \quad \alpha_R(\rho_{\mathcal S}) = \operatorname{Tr}[R(X)\rho_{\mathcal S}].
\label{eq:comparison-rus}
\end{equation}
For a fixed-depth QSP/QSVT implementation, the corresponding expected cost is
\begin{equation}
C_{\mathrm{out}}^{\mathrm{QSVT}} =
\Theta\left( \frac{d}{\widetilde\alpha_d} \right),
\quad
\widetilde\alpha_d = \operatorname{Tr}[\widetilde R_d(X)\rho_{\mathcal S}].
\label{eq:comparison-qsvt}
\end{equation}
A uniform approximation with $\tau=O(\alpha_R)$, with a sufficiently small constant, ensures $\widetilde\alpha_d=\Theta(\alpha_R)$ and constant accuracy of the normalized accepted spectral distribution.
It gives the useful comparison scale
\begin{equation}
C_{\mathrm{out}}^{\mathrm{QSVT}}=\Theta\left(\frac{d}{\alpha_R}\right).
\label{eq:comparison-qsvt-scale}
\end{equation}
For commuting inputs, the same accuracy criterion also controls the trace distance between the corresponding normalized filtered ensembles.

This uniform-approximation benchmark is sufficient for the stated output accuracy; the comparisons below do not assume that every algorithm achieving that accuracy must approximate $R$ uniformly.

\subsection{Regularized resolvent}
\label{sec:resolvent}

Consider a positive semidefinite operator $0\preceq A\preceq\Lambda I_{\mathcal S}$ for some $\Lambda>0$ and the regularized inverse
\begin{equation}
R_\gamma(A)=\gamma(A+\gamma I_{\mathcal S})^{-1}.
\label{eq:resolvent}
\end{equation}
The regularization parameter $\gamma>0$ sets the spectral scale: eigenvalues much smaller than $\gamma$ receive weights close to unity, whereas higher eigenvalues are progressively suppressed.

We also introduce
\begin{equation}
    K=\frac{\Lambda}{\gamma}.
\end{equation}
Using the normalized signal $X=2A/\Lambda-I_{\mathcal S}$ and $\eta=2\gamma/\Lambda$, the target response takes the rational form
\begin{equation}
r_\gamma(x) =
\frac{\eta}{1+x+\eta}.
\label{eq:resolvent-normalized}
\end{equation}
It is a degree-one rational contraction, and therefore admits an exact finite-memory RUS realization.
For an input state $\rho_{\mathcal S}$, the terminal acceptance probability is
\begin{equation}
\begin{aligned}
\alpha_\gamma(\rho_{\mathcal S})
= \operatorname{Tr} [\rho_{\mathcal S}R_\gamma(A)] = \int_0^\Lambda \frac{\gamma}{\lambda+\gamma}d\mu_{\rho_{\mathcal S}}(\lambda).
\end{aligned}
\label{eq:resolvent-weight}
\end{equation}
Conditioning on acceptance reweights the input spectral measure according to
\begin{equation}
\nu_\gamma(B) = \frac{1}{\alpha_\gamma(\rho_{\mathcal S})} \int_B \frac{\gamma}{\lambda+\gamma} d\mu_{\rho_{\mathcal S}}(\lambda).
\label{eq:resolvent-law}
\end{equation}
Thus, accepted spectral measurements preferentially sample low-energy sectors.
When $[\rho_{\mathcal S},A]=0$, the normalized accepted state is
\begin{equation}
\sigma_\gamma = \frac{ \gamma(A+\gamma I_{\mathcal S})^{-1} \rho_{\mathcal S} }{\alpha_\gamma(\rho_{\mathcal S})}.
\label{eq:resolvent-ensemble}
\end{equation}

The filter admits a particularly simple one-query renewal compilation.
Choosing the constant multiplier $S=1/(2+\eta)$ gives
\begin{equation}
\begin{aligned}
p_{\mathsf a}(x) =\frac{\eta}{2+\eta}, &\quad
p_{\mathsf r}(x) =\frac{1+x}{2+\eta}, \\
p_{\mathsf c}(x) &=\frac{1-x}{2+\eta}.
\end{aligned}
\label{eq:res-probs}
\end{equation}
Equivalently, in the original spectral variable $\lambda$,
\begin{equation}
\begin{aligned}
p_{\mathsf a}(\lambda) =\frac{\gamma}{\Lambda+\gamma},&\quad
p_{\mathsf r}(\lambda) =\frac{\lambda}{\Lambda+\gamma},\\
p_{\mathsf c}(\lambda) &=\frac{\Lambda-\lambda}{\Lambda+\gamma}.
\end{aligned}
\label{eq:res-probs-physical}
\end{equation}
An explicit set of causal spectral factors is
\begin{equation}
\begin{aligned}
g_{\mathsf a}(z)=\sqrt{\frac{\eta}{2+\eta}},&\quad
g_{\mathsf r}(z)=\frac{1+z}{\sqrt{2(2+\eta)}},\\
g_{\mathsf c}(z)&=\frac{1-z}{\sqrt{2(2+\eta)}}.
\end{aligned}
\label{eq:res-factors}
\end{equation}
Their squared magnitudes reproduce the three trial probabilities, and the degree-stripping compiler implements the trial with one controlled-walk query.

A notable feature is that the probability of acceptance in each elementary trial, $p_{\mathsf a}=\gamma/(\Lambda+\gamma)$, is independent of the eigenvalue.
Consequently, the expected total number of queries until acceptance, including fresh inputs after rejection, is exactly
\begin{equation}
C_{\mathrm{out}}^{\mathrm{RUS}}= \frac{1}{p_{\mathsf a}} = \frac{\Lambda+\gamma}{\gamma} = K+1.
\label{eq:res-cost}
\end{equation}
This result holds for every input spectral distribution. 
Appendix~\ref{app:resolvent-cost} provides a detailed derivation of the exact query cost, establishes a minimax property within the considered compiler family, and develops a statistical estimator for the resolvent spectral weight.

The length of an individual run depends on the input spectrum, but the total expected query count per accepted output does not.
We compare this exact result with fixed-depth polynomial postselection.
The target response can be written as
\begin{equation}
r_K(x) = \frac{2/K}{1+x+2/K}.
\label{eq:qK}
\end{equation}
The required polynomial degree follows from an exact minimax bound.
For $f_s(t)=s/(t+s)$ on $[0,1]$, the best degree-$D$ uniform polynomial approximation error is $E_D(f_s)=\exp[-2D\operatorname{arsinh}\sqrt{s}]/[2(1+s)]$, obtained from the classical absolute-error minimax formula for $1/t$~\cite{PlonkaTasche2026}.
Since $r_K(x)=f_{1/K}((1+x)/2)$, this gives
\begin{equation}
E_D(r_K)=\frac{\exp[-2D\operatorname{arsinh}(K^{-1/2})]}{2(1+K^{-1})}.
\label{eq:res-minimax-error}
\end{equation}
Therefore the minimum degree is $\Theta(\sqrt K\log(1/\tau))$ for large $K$ and small $\tau$.
The query-degree correspondence established in Sec.~\ref{sec:comparison-model} then implies
\begin{equation}
d_{\mathrm{res}}^{\mathrm{QSVT}} = \Omega\left(\sqrt K\log\frac1\tau\right).
\label{eq:res-qsvt-degree}
\end{equation}
To evaluate the resulting end-to-end cost, consider a broad input spectrum with a uniform spectral distribution on $[0,\Lambda]$.
Its acceptance probability is
\begin{equation}
\alpha_\gamma = \frac1\Lambda \int_0^\Lambda \frac{\gamma}{\lambda+\gamma} d\lambda= \frac{\log(1+K)}{K}.
\label{eq:res-broad-alpha}
\end{equation}
The same scaling holds for sufficiently regular spectral densities that remain positive near the lower spectral endpoint.

Choosing uniform approximation error $\tau=O(\alpha_\gamma)$ guarantees constant accuracy of the normalized spectral distribution, and of the filtered ensemble for commuting inputs.
Under this approximation requirement,
\begin{equation}
\begin{aligned}
C_{\mathrm{out}}^{\mathrm{RUS}} =\Theta(K),\quad C_{\mathrm{out}}^{\mathrm{QSVT}}=\Omega(K^{3/2}).
\end{aligned}
\label{eq:res-separation}
\end{equation}

Thus, the exact rational realization achieves a polynomial reduction in expected accepted-output queries relative to the specified uniformly approximating fixed-depth polynomial baseline.

\subsection{Lorentzian spectroscopy}
\label{sec:lorentz}
We next consider selecting spectral components near a target energy.
Let $X=X^\dagger$ have spectrum in $[-1,1]$ and define the Lorentzian filter
\begin{equation}
R_{\xi,\eta}(x)
=\frac{\eta^2}{(x-\xi)^2+\eta^2},
\label{eq:lorentz}
\end{equation}
where $\xi\in[-1,1]$ specifies the center and $\eta>0$ is the half-width at half-maximum.

The filter assigns unit weight at $x=\xi$ and suppresses components farther from the center.
For an input state $\rho_{\mathcal S}$, the acceptance probability is
\begin{equation}
\alpha_{\xi,\eta}(\rho_{\mathcal S})
=\int_{-1}^{1}
\frac{\eta^2}{(x-\xi)^2+\eta^2}
\,d\mu_{\rho_{\mathcal S}}(x).
\label{eq:lorentz-weight}
\end{equation}
Thus, acceptance measures the spectral weight near $\xi$, with contributions weighted by the Lorentzian profile.

To express this in physical energy units, let $H=\alpha X$, where $\alpha>0$, and write $\xi=\omega/\alpha$ and $\eta=\Delta/\alpha$.
Here, $\omega$ is the target energy and $\Delta$ is the energy half-width.
The acceptance probability then satisfies
\begin{equation}
\frac{\alpha_{\xi,\eta}(\rho_{\mathcal S})}{\pi\Delta}
=-\frac{1}{\pi}\operatorname{Im}
\operatorname{Tr}\left[
\rho_{\mathcal S}
(\omega+i\Delta-H)^{-1}
\right].
\label{eq:spectroscopy}
\end{equation}
Scanning $\omega$ at fixed $\Delta$ therefore maps out the input state's spectral distribution broadened by a Lorentzian.
Alternatively, measuring the energy of accepted outputs yields samples from this distribution reweighted toward the selected energy window.

If the input state commutes with $X$, the accepted state is obtained by reweighting its spectral populations with the Lorentzian filter:
\begin{equation}
\sigma_{\xi,\eta}
=\frac{R_{\xi,\eta}(X)\rho_{\mathcal S}}
{\alpha_{\xi,\eta}(\rho_{\mathcal S})}.
\label{eq:lorentz-ensemble}
\end{equation}
The Lorentzian filter has rational degree two and can be realized exactly by repeating a trial with two controlled-walk queries.
To construct this trial, define
\begin{equation}
b=1+|\xi|,
\qquad
M=b^2+\eta^2,
\label{eq:lorentz-scale}
\end{equation}
so that $|x-\xi|\leq b$ throughout $[-1,1]$.
Choosing the constant multiplier $S=1/M$ gives the trial probabilities
\begin{equation}
\begin{aligned}
p_{\mathsf a}(x)&=\frac{\eta^2}{M},\\
p_{\mathsf r}(x)&=\frac{(x-\xi)^2}{M},\\
p_{\mathsf c}(x)&=\frac{b^2-(x-\xi)^2}{M}.
\end{aligned}
\label{eq:lorentz-trial}
\end{equation}
These probabilities are nonnegative, sum to one, and have degree at most two.

The acceptance probability is constant within a single trial; the Lorentzian dependence emerges from repeating the trial after continuation.
To implement these probabilities, we construct polynomial amplitudes satisfying $|g_j(z)|^2=p_j((z+z^{-1})/2)$ on the unit circle.
For acceptance and rejection, we may choose
\begin{equation}
\begin{aligned}
g_{\mathsf a}(z)&=\frac{\eta}{\sqrt M},\\
g_{\mathsf r}(z)&=\frac{1-2\xi z+z^2}{2\sqrt M}.
\end{aligned}
\label{eq:lorentz-factors}
\end{equation}
Fej\'er--Riesz factorization supplies a continuation amplitude $g_{\mathsf c}(z)$ of degree at most two. Explicit polynomial spectral factors are derived in Appendix~\ref{app:lorentz}.

All three amplitudes contain only nonnegative powers of $z$ and form a column of unit norm on the unit circle.
The compilation procedure therefore implements the complete trial using two forward controlled-walk queries.
As in the resolvent example, the one-trial acceptance probability is independent of the spectral value.
The expected total query cost per accepted output is therefore exactly
\begin{equation}
C_{\mathrm{out}}^{\mathrm{RUS}} = \frac{2}{p_{\mathsf a}} = \frac{2M}{\eta^2} =\Theta(\eta^{-2})
\label{eq:lorentz-cost}
\end{equation}
for fixed $\xi$ as $\eta\to0$.
The exact cost is independent of the input spectral distribution.

For comparison, the optimal polynomial approximation degree can be bounded using the same reciprocal minimax formula.
For a fixed interior center $|\xi|<1$, let $h=1-|\xi|>0$, $b=1+|\xi|$, and $m=\lfloor D/2\rfloor$.
Restricting the approximation interval to $[\xi-h,\xi+h]$ and symmetrizing any degree-$D$ approximant about $\xi$ yields a degree-$m$ polynomial in $(x-\xi)^2$; this gives a lower bound. 
Approximating in $(x-\xi)^2$ over its full range $[0,b^2]$ gives an upper bound.
In terms of the minimax error $E_D$ and the function $f_s$ defined above, the resulting bounds are
\begin{equation}
E_m(f_{(\eta/h)^2})\leq E_D(R_{\xi,\eta})
\leq E_m(f_{(\eta/b)^2}).
\label{eq:lorentz-minimax-bounds}
\end{equation}
Since $h$ and $b$ are positive constants for fixed interior $\xi$, these bounds imply that the minimum degree necessary and sufficient for uniform error $\tau$ is $\Theta(\eta^{-1}\log(1/\tau))$ as $\eta,\tau\to0$.
The query-degree correspondence from Sec.~\ref{sec:comparison-model} therefore gives
\begin{equation}
d_{\mathrm{Lor}}^{\mathrm{QSVT}} =
\Theta\left(
\eta^{-1}\log\frac1\tau
\right)
\label{eq:lorentz-qsvt-degree}
\end{equation}
with constants depending on the fixed interior center $\xi$.

Consider a broad spectral measure with a smooth density $w(x)$, continuous and strictly positive at an interior center $\xi$.
Then
\begin{equation}
\begin{aligned}
\alpha_{\xi,\eta} &= \int_{-1}^{1} \frac{\eta^2 w(x)} {(x-\xi)^2+\eta^2}\,dx\\
&= \pi\eta w(\xi)+o(\eta) =\Theta(\eta).
\end{aligned}
\label{eq:lorentz-broad-alpha}
\end{equation}
Taking uniform approximation error $\tau=O(\eta)$ is sufficient for constant accuracy of the normalized accepted spectrum.

Combining the approximation degree with the acceptance probability gives
\begin{equation}
\begin{aligned}
C_{\mathrm{out}}^{\mathrm{RUS}}=\Theta(\eta^{-2}),\quad C_{\mathrm{out}}^{\mathrm{QSVT}}=\Theta\left(\eta^{-2}\log\eta^{-1} \right).
\end{aligned}
\label{eq:lorentz-separation}
\end{equation}
In this comparison, the RUS circuit avoids the logarithmic factor required by the uniformly approximating polynomial baseline.

\subsection{Query cost per accepted output}
\label{sec:application-table}
We bring together the two examples by comparing the expected number of controlled-walk queries needed to obtain one accepted output, including continuation trials and independent restarts after rejection.
The RUS constructions implement the target responses exactly, using one query per trial for the resolvent and two for the Lorentzian.
Their constant one-trial acceptance probabilities make the expected cost per accepted output independent of the input spectral distribution.

Figure~\ref{fig:application-filters} compares the target rational filters, the polynomial branch amplitudes implementing their trials, and the corresponding nontermination probabilities as functions of the number of trials.
The numerical parameters and procedures underlying these illustrations are specified in Appendix~\ref{app:numerical_experiments}.
Table~\ref{tab:application-complexity} summarizes the resulting query costs and compares them with the fixed-depth polynomial baselines in the representative spectral regimes analyzed above.
\begin{table}[t]
\centering
\caption{
Expected controlled-walk queries per accepted output.
The QSP/QSVT baselines use fixed-depth polynomial postselection and independent restarts, without amplitude amplification.
The uniform acceptance-response error is chosen as $\tau=c\alpha_R$, with a sufficiently small fixed constant $c>0$, to ensure fixed accuracy of the normalized accepted spectral distribution.
The resolvent regime has $K=\Lambda/\gamma$ and $\alpha_\gamma=\Theta(\log K/K)$; the Lorentzian regime has an interior spectral center and $\alpha_{\xi,\eta}=\Theta(\eta)$.
}
\label{tab:application-complexity}
\renewcommand{\arraystretch}{1.3}
\setlength{\tabcolsep}{14pt}
\begin{tabular}{@{}lcc@{}}
\toprule
Task
& $C_{\mathrm{out}}^{\mathrm{RUS}}$
& $C_{\mathrm{out}}^{\mathrm{QSVT}}$ \\
\midrule
Resolvent
& $\Theta(K)$
& $\Omega(K^{3/2})$ \\
Lorentzian
& $\Theta(\eta^{-2})$
& $\Theta(\eta^{-2}\log\eta^{-1})$ \\
\bottomrule
\end{tabular}
\end{table}

For the resolvent, Table~\ref{tab:application-complexity} establishes a polynomial reduction in query cost relative to the stated baseline.
For the Lorentzian, the exact RUS construction avoids the logarithmic approximation overhead in the polynomial construction analyzed above.
Both comparisons account for the repetitions needed to obtain an accepted output, rather than only the cost of a single trial or circuit execution.

\section{Conclusion}
\label{sec:discussion}

We have developed a finite-memory, measured repeat-until-success framework for rational spectral filtering of Hermitian operators.
Its central mechanism is the accumulation of terminal probabilities over first-hitting histories.
Although the protocol may run for arbitrarily many rounds, these contributions can be summed through a finite-dimensional Lyapunov equation.
Under uniform termination and phase-inversion symmetry, the realizable spectral effects are precisely the real rational functions with no poles on $[-1,1]$ and values in $[0,1]$.
Our constructive compiler realizes these functions by converting polynomial trial probabilities into quantum amplitudes, decomposing them into forward controlled-walk queries and signal-independent unitaries, and using a clock register to repeat the resulting trial.
The construction provides explicit bounds on controller memory, termination time, and expected query cost.

The framework also describes the complete terminal quantum operations, including their action on coherences between eigenspaces.
A positive-semidefinite coherence kernel captures this action: its diagonal determines the spectral weights, while its off-diagonal values describe the transformation of coherences.
A rank-one kernel corresponds to a single-Kraus operation.
This distinction shows that realizing a target acceptance effect and controlling the coherence of the accepted state are separate aspects of filter design.

Applications to regularized resolvents and Lorentzian spectroscopy yield exact realizations with one and two controlled-walk queries per trial, respectively.
Their expected costs per accepted output can be calculated explicitly, including continuation trials and restarts after rejection.
The comparisons identify query-cost advantages relative to the fixed-depth polynomial-postselection baselines analyzed here.

These results provide a constructive connection between rational spectral functions and quantum protocols with finite memory and intermediate measurements.
Future directions include reducing memory requirements, compiling filters with prescribed coherence properties, and extending the construction to multiple terminal channels.

\section*{Acknowledgments}
The authors acknowledge support by the Federal Academic Leadership Program Priority 2030 (NUST MISIS Strategic Technology Project "Quantum Internet").


\appendix

\section{Necessity and the memory--degree bound}
\label{app:necessity}

\subsection{First-hitting normalization and uniform stability}

We use the controller notation fixed in the main text: $\mathsf C$ is the continuation block and $\mathsf T_j$ are the terminal blocks. Their routed counterparts and the controller input embedding are
\begin{equation}
 \begin{aligned}
 \mathsf C_W=(\mathsf C\otimes I_{\mathcal W})\Gamma_W,\quad \mathsf T_{j,W}=(\mathsf T_j\otimes I_{\mathcal W})\Gamma_W.
 \end{aligned}
 \label{eq:joint-branches}
\end{equation}
The first-hitting operators are
\begin{equation}
 K_{j,n}=\mathsf T_{j,W}\mathsf C_W^nV_\beta,
 \quad j\in\{\mathsf a,\mathsf r\},\quad n\geq0.
 \label{eq:appendix-first-hitting}
\end{equation}
Unitarity of the colligation and of the router gives the one-round completeness identity
\begin{equation}
 \mathsf C_W^\dagger\mathsf C_W+\sum_{j\in\{\mathsf a,\mathsf r\}}\mathsf T_{j,W}^\dagger\mathsf T_{j,W}=I_{\mathcal C}\otimes I_{\mathcal W}.
 \label{eq:branch-complete}
\end{equation}
Multiplying by continuation powers on both sides and summing produces the finite telescoping identity
\begin{equation}
 \begin{aligned}
 \sum_{j\in\{\mathsf a,\mathsf r\}}\sum_{n=0}^{N-1}K_{j,n}^\dagger K_{j,n}+V_\beta^\dagger(\mathsf C_W^\dagger)^N\mathsf C_W^NV_\beta
 =I_{\mathcal W}.
 \end{aligned}
 \label{eq:finite-complete}
\end{equation}
The remainder is the POVM effect of $N$ consecutive continuation outcomes. 
It vanishes in operator norm under uniform termination, which yields $F_{\mathsf a}+F_{\mathsf r}=I_{\mathcal W}$.

For completeness, uniform stability implies a geometric estimate even when $\|\mathsf C(z)\|=1$, as may occur in a clocked realization.
Choose $N_0$ such that
\begin{equation}
 q_0:=\sup_{z\in\mathbb T}\|\mathsf C(z)^{N_0}\|<1.
 \label{eq:uniform-block-contraction}
\end{equation}
All one-round continuation blocks are contractions. 
For $n=kN_0+s$, $0\leq s<N_0$, submultiplicativity therefore gives
\begin{equation}
 \sup_{z\in\mathbb T}\|\mathsf C(z)^n\|\leq q_0^k.
 \label{eq:uniform-geometric}
\end{equation}
This bound proves uniform convergence of the Lyapunov series and of the mixed series used for quantum channels. 
It also gives uniqueness:
if $\Delta=\mathsf C(z)^\dagger\Delta\mathsf C(z)$, then
$\Delta=\mathsf C(z)^{\dagger n}\Delta\mathsf C(z)^n$ for every
$n$, and the right-hand side tends to zero.

\subsection{Rationality and phase-inversion descent}

Let $d=\dim\mathcal H_{\mathcal C}$ and introduce
\begin{equation}
 C_0=\mathsf C(I_{\mathcal C}-\Pi_s),~  C_1=\mathsf C\Pi_s,~ \mathsf C(z)=C_0+zC_1.
 \label{eq:appendix-continuation-affine}
\end{equation}
The column-vectorization identity $\operatorname{vec}(AYB)=(B^{\mathsf T}\otimes A) \operatorname{vec}Y$ turns the Lyapunov equation into a finite linear system with coefficient matrix
\begin{equation}
 \mathsf L(z)=I_{d^2}-
 \mathsf C(z)^{\mathsf T}\otimes\mathsf C(z)^\dagger.
 \label{eq:Lyapunov-matrix}
\end{equation}
On $\mathbb T$, its entries and those of $S_{\mathsf a}(z)=\mathsf T_{\mathsf a}(z)^\dagger \mathsf T_{\mathsf a}(z)$ are Laurent polynomials supported in $\{-1,0,1\}$. 
The spectral-radius condition $\rho(\mathsf C(z))<1$ ensures that $\mathsf L(z)$ is invertible.
By the adjugate formula,
\begin{equation}
 \operatorname{vec}Y_{\mathsf a}(z)
 =\frac{\operatorname{adj}\mathsf L(z)}{\det\mathsf L(z)}
 \operatorname{vec}S_{\mathsf a}(z).
 \label{eq:adjugate}
\end{equation}
Taking the expectation in $\ket\beta$ proves that $f_{\mathsf a}(z)$ is a rational function with no poles on the physical unit circle. 
Apparent singularities in a nonreduced representation are removable there.

To express the response as a function of the Hermitian eigenvalue, we use the identity
\begin{equation}
 f(z)=f(z^{-1})\Longleftrightarrow
 f(z)=R\left(\frac{z+z^{-1}}2\right)
  \label{eq:fixed-field}
\end{equation}
for some $R\in\mathbb C(x)$.
To prove it, put $t=z+z^{-1}$. 
The relation $z^2-tz+1=0$ makes $\mathbb C(z)$ a quadratic extension of $\mathbb C(t)$. 
Each rational function of $z$ has a unique expression $a(t)+zb(t)$ with $a,b\in\mathbb C(t)$. 
Under $z\mapsto z^{-1}=t-z$, this becomes $a(t)+tb(t)-zb(t)$, so invariance forces $b=0$.
The descended function is unique because $x(z)=t/2$ is surjective onto $[-1,1]$ on the unit circle. 
An identity holding for all $z\in\mathbb T$ extends to an identity of rational functions.

For an exact RUS realization, $f_{\mathsf a}(z)$ is real and lies in $[0,1]$ on $\mathbb T$. 
The descended $R(x)$ therefore takes real values in $[0,1]$ for every $x\in[-1,1]$, and comparison with its coefficient-conjugate rational function shows $R\in\mathbb R(x)$. 
If the reduced $R$ had a pole on the signal interval, composition with $x(z)$ would yield a singularity on $\mathbb T$, including at the branch points $x=\pm1$. 
This contradicts the regularity of the phase response. 
Finally, a coprime real denominator has a constant sign on the connected interval; choosing it positive yields $Q>0$ and $0\leq P\leq Q$.

\subsection{Rational degree and controller dimension}
Put $m=d^2$. 
Since $\mathsf L(z)$ is an $m\times m$ Laurent matrix with entries supported in $[-1,1]$, its determinant has support in $[-m,m]$. Every entry of its adjugate has support in $[-(m-1),m-1]$, and multiplication by the source enlarges this range by at most one. 
Consequently, the numerator and denominator of $f_{\mathsf a}$ can both be represented by Laurent polynomials supported in $[-m,m]$. Multiplication by $z^m$ produces ordinary polynomials of degree at most $2m$; cancellation cannot increase that degree.

Let $R$ be nonconstant with reduced rational degree $r=\max(\deg P,\deg Q)$. 
The rational map $x(z)=(z+z^{-1})/2$ has degree two, and degrees of nonconstant rational maps multiply under composition. 
Hence
\begin{equation}
 \deg_z R(x(z))=2r.
 \label{eq:degree-composition}
\end{equation}
Since this map equals $f_{\mathsf a}(z)$, whose rational degree is at most $2d^2$, we obtain
\begin{equation}
 r\leq d^2.
 \label{eq:appendix-memory-degree}
\end{equation}
This proves the necessity and memory-degree statements of Theorem~\ref{thm:representation}.

\section{Constructive sufficiency and polynomial compilation}
\label{app:compiler}

\subsection{Polynomial trial probabilities and causal factors}

Let $R=P/Q$ satisfy Eq.~\eqref{eq:representation}. 
The admissible multiplier $S$ and the probabilities $p_j$ have already been defined in Eqs.~\eqref{eq:multiplier-condition} and \eqref{eq:representation-trial}.
For the baseline choice $S=1/Q_{\max}$, every trial probability has degree at most $r=\operatorname{rdeg}R$ and $1-p_{\mathsf c}(x)\geq Q_{\min}/Q_{\max}>0$.
We describe the spectral factorization for an arbitrary trial degree $L$.

Write $p_j(x)=\sum_{q=0}^{L}p_{j,q}x^q$. 
Its phase lift is
\begin{equation}
 \widehat p_j(z)=p_j\left(\frac{z+z^{-1}}2\right)
 =\sum_{\ell=-L}^{L}c_{j,\ell}z^\ell,
 \label{eq:lift}
\end{equation}
with explicit coefficients
\begin{equation}
 c_{j,\ell}=
 \sum_{\substack{|\ell|\leq q\leq L\\q-\ell\ \mathrm{even}}}
 2^{-q}\binom{q}{(q-\ell)/2}p_{j,q}.
 \label{eq:lift-coeff}
\end{equation}
Because $p_j\geq0$ on $[-1,1]$, the Laurent polynomial $\widehat p_j$ is nonnegative on $\mathbb T$. 
Define the reciprocal-conjugate operation
\begin{equation}
 g^\sharp(z):=\overline{g(1/\overline z)}.
 \label{eq:sharp}
\end{equation}
The Fej\'er--Riesz theorem supplies a causal polynomial $g_j(z)=\sum_{k=0}^L\gamma_{j,k}z^k$ satisfying
\begin{equation}
 \widehat p_j(z)=g_j^\sharp(z)g_j(z),
 \qquad \sum_jg_j^\sharp(z)g_j(z)=1
 \quad (z\in\mathbb T).
 \label{eq:appendix-spectral-factors}
\end{equation}
For $0\leq\ell\leq L$, this is equivalent to the finite autocorrelation equations
\begin{equation}
 c_{j,\ell}=\sum_{k=0}^{L-\ell} \overline{\gamma_{j,k}}\gamma_{j,k+\ell},\quad c_{j,-\ell}=\overline{c_{j,\ell}}.
 \label{eq:autocorrelation}
\end{equation}

The factor may also be obtained algebraically from polynomial roots.
For a nonzero $\widehat p$ of actual Laurent degree $l$, the polynomial $z^l\widehat p(z)$ has reciprocal-conjugate root pairs $\{\zeta,1/\overline\zeta\}$. 
Roots on the unit circle have even multiplicity because $\widehat p$ is nonnegative there.
Selecting one root from each off-circle pair and half of each unit-circle multiplicity produces a degree-$l$ causal factor, after fixing a scalar normalization. 
A zero branch is assigned $g_j=0$, and a nonnegative constant branch is assigned its nonnegative square root. 
This gives an exact algebraic construction. It does not, by itself, bound the numerical cost or conditioning of the factorization.

\subsection{Degree stripping}

Let $\ket{g^{(L)}(z)}=\sum_{k=0}^Lz^k\ket{g_k^{(L)}}$ denote the normalized polynomial column of Eq.~\eqref{eq:inner-column}. 
For each nonzero degree $\ell=L,L-1,\ldots,1$, define
\begin{equation}
 \Lambda_\ell=
 \frac{\ket{g_\ell^{(\ell)}}\bra{g_\ell^{(\ell)}}}
 {\langle g_\ell^{(\ell)}|g_\ell^{(\ell)}\rangle},
 \label{eq:strip-projector}
\end{equation}
and apply the inverse elementary factor
\begin{equation}
 \ket{g^{(\ell-1)}(z)}=
 (I_3-\Lambda_\ell+z^{-1}\Lambda_\ell)
 \ket{g^{(\ell)}(z)}.
 \label{eq:strip-step}
\end{equation}
The inner identity forces the coefficient of $z^\ell$ in $\langle g^{(\ell)}(z)|g^{(\ell)}(z)\rangle$ to vanish, giving $\langle g_0^{(\ell)}|g_\ell^{(\ell)}\rangle=0$.
Thus $\Lambda_\ell\ket{g_0^{(\ell)}}=0$, eliminating the potential $z^{-1}$ term. 
By definition, $(I_3-\Lambda_\ell)\ket{g_\ell^{(\ell)}}=0$, eliminating the highest positive power. 
The remaining coefficients obey
\begin{equation}
 \ket{g_k^{(\ell-1)}}=
 (I_3-\Lambda_\ell)\ket{g_k^{(\ell)}}+
 \Lambda_\ell\ket{g_{k+1}^{(\ell)}},
 \quad 0\leq k<\ell.
 \label{eq:strip-recursion}
\end{equation}
The inverse factor is unitary on $\mathbb T$, so the reduced column remains inner. 
Repetition produces a constant unit vector $\ket u$, and reversing the recursion proves Eq.~\eqref{eq:factorization}. 
If the leading coefficient is zero, we first replace $\ell$ by the actual degree.

For each rank-one projector $\Lambda_\ell$, choose a unitary $U_\ell$ such that $\Lambda_\ell=U_\ell\ket s\bra s U_\ell^\dagger$.
Its remaining columns may be any orthonormal completion.
Equation~\eqref{eq:section} then realizes the corresponding forward factor using exactly one controlled-$W$ query.
The complete trial operators are
\begin{equation}
 \begin{aligned}
 G_j(W)
 &= (\bra j\otimes I_{\mathcal W})
 E_{\Lambda_L}(W)\cdots E_{\Lambda_1}(W)
 (\ket u\otimes I_{\mathcal W})\\
 &=g_j(W),
 \end{aligned}
 \label{eq:trial-Kraus}
\end{equation}
where $j\in\{\mathsf a,\mathsf r,\mathsf c\}$. 
Spectral calculus yields $G_j^\dagger G_j=p_j(X_{\mathcal W})$ and $\sum_jG_j^\dagger G_j=I_{\mathcal W}$. 
Moreover, $\|G_{\mathsf c}\|^2=\max_xp_{\mathsf c}(x)<1$, uniformly over the physical signal interval. 
The geometric renewal series therefore converges in operator norm and gives Eq.~\eqref{eq:compiled-effect}.
This proves exact sufficiency at the level of complete trials.

\section{Fixed clocked colligation}
\label{app:clock}

The factorization above specifies a scheduled sequence of $L$ one-query sections. 
We now implement it with one fixed controller, one fixed router, and one flag measurement after each elementary round. 
For $L\geq1$, use the controller memory
\begin{equation}
 \begin{aligned}
 \mathcal H_{\mathcal C}^{\mathrm{clk}}=\mathbb C^L\otimes\mathbb C^3,~ \ket{\beta_{\mathrm{clk}}}=\ket1\otimes U_1^\dagger\ket u,~ \Pi_s^{\mathrm{clk}} =I_L\otimes\ket s\bra s.
 \end{aligned}
 \label{eq:clock-data}
\end{equation}
Let $\ket\ell$ denote the clock basis. 
At the final clock step, define the three row operators, by
\begin{equation}
 R_j=(\bra L\otimes\bra j U_L),
 \qquad j\in\{\mathsf a,\mathsf r,\mathsf c\}.
 \label{eq:clock-final-rows}
\end{equation}
The signal-independent continuation and terminal blocks are
\begin{align}
 \mathsf C^{\mathrm{clk}}
 &=\sum_{\ell=1}^{L-1}
 \ket{\ell+1}\bra\ell\otimes U_{\ell+1}^\dagger U_\ell
 +\ket{\beta_{\mathrm{clk}}}R_{\mathsf c},
 \label{eq:clock-A}\\
 \mathsf T_j^{\mathrm{clk}}
 &=\ket{\beta_{\mathrm{clk}}}R_j,
 \qquad j\in\{\mathsf a,\mathsf r\}.
 \label{eq:clock-C}
\end{align}
These fixed blocks implement a basis change and a clock advance at intermediate sections. 
At the final section, they reset the clock and write the branch outcome into the measured flag.

The advancing terms have mutually orthogonal clock ranges $2,\ldots,L$, orthogonal to the reset range at clock value one.
Moreover, the three final rows $\bra j U_L$ form an orthonormal basis of the trial branch space. 
Therefore
\begin{equation}
 (\mathsf C^{\mathrm{clk}})^\dagger\mathsf C^{\mathrm{clk}}
 +\sum_{j\in\{\mathsf a,\mathsf r\}}
 (\mathsf T_j^{\mathrm{clk}})^\dagger\mathsf T_j^{\mathrm{clk}}
 =I_{3L}.
 \label{eq:clock-isometry}
\end{equation}
Stacking the three blocks gives an isometry from the continuing-flag input sector into the full flag-controller space. 
An orthonormal completion of this isometry produces the fixed signal-independent unitary colligation $\mathsf U^{\mathrm{clk}}$ acting on a $9L$-dimensional flag-controller space; its unused input columns do not participate in the measured protocol.

For a phase $z$, introduce
$\Gamma_{\mathrm{clk}}(z)= I_L\otimes(I_3-\ket s\bra s+z\ket s\bra s)$ and the routed blocks
$\mathsf C_{\mathrm{clk}}(z)=\mathsf C^{\mathrm{clk}}
\Gamma_{\mathrm{clk}}(z)$ and
$\mathsf T_{j,\mathrm{clk}}(z)=\mathsf T_j^{\mathrm{clk}}
\Gamma_{\mathrm{clk}}(z)$.
The successive basis changes telescope, giving
\begin{align}
 \mathsf T_{j,\mathrm{clk}}(z)
 \mathsf C_{\mathrm{clk}}(z)^{L-1}
 \ket{\beta_{\mathrm{clk}}}
 &=g_j(z)\ket{\beta_{\mathrm{clk}}},
 \label{eq:clock-terminal}\\
 \mathsf C_{\mathrm{clk}}(z)^L
 \ket{\beta_{\mathrm{clk}}}
 &=g_{\mathsf c}(z)\ket{\beta_{\mathrm{clk}}}.
 \label{eq:clock-return}
\end{align}
The measured flag returns continuation deterministically during the first $L-1$ rounds. 
After $n$ complete continuing trials, a terminal outcome therefore carries amplitude $g_j(z)g_{\mathsf c}(z)^n$, matching the scheduled circuit exactly.

It remains to establish uniform termination on the \emph{entire} clocked memory space, not only on the chosen initial state.
The operator $\mathsf C_{\mathrm{clk}}(z)^L$ is block diagonal in the clock register: in each clock sector, its block $B_\ell(z)$ has rank at most one and norm at most one. 
The cyclic invariance of the trace of the ordered section product gives $\operatorname{Tr}B_\ell(z)=g_{\mathsf c}(z)$ for every $\ell$.
Every rank-one matrix satisfies $B_\ell(z)^k=(\operatorname{Tr}B_\ell(z))^{k-1}B_\ell(z)$ for $k\geq1$. Hence, for $n=kL+s$, $k\geq1$, $0\leq s<L$,
\begin{equation}
 \sup_{z\in\mathbb T}\|\mathsf C_{\mathrm{clk}}(z)^n\|
 \leq(1-\tau_{\min})^{(k-1)/2}
 \longrightarrow0.
 \label{eq:clock-stability}
\end{equation}
Here $\tau_{\min}=\min_x S(x)Q(x)>0$. 
If $\tau_{\min}=1$, each block $B_\ell(z)$ is nilpotent of order at most two, and termination occurs after finitely many rounds. This proves uniform termination for the full fixed controller.

For the baseline multiplier $S=1/Q_{\max}$, the scheduled length is $L\leq r$ and the active controller dimension is $3L\leq3r$.
The construction therefore uses $\lceil\log_2(3L)\rceil$ memory qubits under a general binary embedding, or $\lceil\log_2L\rceil+2$ memory qubits under a clock--branch product embedding. 
Extra unused basis states may be assigned immediate rejection with inactive routing, preventing spurious nonterminating sectors. 
The termination and expected query bounds are those derived in Sec.~\ref{subsec:cost}.

For a constant target $R(x)=a\in[0,1]$, a one-dimensional memory suffices: take $\Pi_s=0$, $\mathsf C=0$, $\mathsf T_{\mathsf a}=\sqrt a$, and $\mathsf T_{\mathsf r}=\sqrt{1-a}$.
The corresponding colligation is obtained by isometric completion and uses no walk queries.

\section{Terminal operations and coherence kernels}
\label{app:channel}

Let $W=\sum_\nu z_\nu\Omega_\nu$ and recall the first-hitting amplitudes
\begin{equation}
 \ket{k_{j,n}(z)}
 =\mathsf T_j(z)\mathsf C(z)^n\ket\beta.
 \label{eq:history-vector}
\end{equation}
For a walk eigenvector $\ket{\omega_z}$, we have $K_{j,n}\ket{\omega_z}= \ket{k_{j,n}(z)}\otimes\ket{\omega_z}$.
Taking the controller trace in the terminal operation yields
\begin{equation}
 \Phi_j(\rho_{\mathcal W})  =\sum_{\nu,\mu}\chi_j(z_\nu,z_\mu) \Omega_\nu\rho_{\mathcal W}\Omega_\mu,
 \label{eq:appendix-walk-channel}
\end{equation}
where $\chi_j(z,w)=\sum_{n\geq0} \langle k_{j,n}(w)|k_{j,n}(z)\rangle$.
The sum is incoherent across distinct stopping times.
For a finite collection of phases, the kernel matrix is a Gram matrix; in particular,
\begin{equation}
 \chi_j(z,z)=f_j(z),\qquad
 |\chi_j(z,w)|^2\leq f_j(z)f_j(w).
 \label{eq:kernel-bound}
\end{equation}

For the mixed Lyapunov representation, define $S_j(z,w)=\mathsf T_j(w)^\dagger\mathsf T_j(z)$ and
\begin{equation}
 Y_j(z,w)=\sum_{n\geq0}
 \mathsf C(w)^{\dagger n}S_j(z,w)\mathsf C(z)^n.
 \label{eq:mixed-series}
\end{equation}
The series is uniformly convergent by Eq.~\eqref{eq:uniform-geometric} and satisfies
\begin{align}
 Y_j(z,w)&=S_j(z,w)+
 \mathsf C(w)^\dagger Y_j(z,w)\mathsf C(z),
 \label{eq:mixed-Lyapunov}\\
 \chi_j(z,w)&=\bra\beta Y_j(z,w)\ket\beta.
 \label{eq:mixed-kernel}
\end{align}
The difference of two solutions obeys the homogeneous mixed equation and vanishes upon iteration, proving uniqueness. 
Vectorization gives
\begin{equation}
 \begin{aligned}
 \operatorname{vec}Y_j(z,w)
 ={}&[I_{d^2}-\mathsf C(z)^{\mathsf T}\otimes
 \mathsf C(w)^\dagger]^{-1}\\
 &\times\operatorname{vec}S_j(z,w).
 \end{aligned}
 \label{eq:mixed-vector}
\end{equation}
For $z,w\in\mathbb T$, the entries of this finite resolvent are rational in $z$ and $w^{-1}$. 
The diagonal $z=w$ recovers the ordinary Lyapunov equation for $Y_j(z)$.

To compress the operation to the original system, use the form of the specturm of $W$ and the decomposition of the
initial block-encoding qubit into $\ket{v_+}$ and $\ket{v_-}$.
The partial trace over that qubit retains only terms with equal ancilla labels. 
This gives
\begin{equation}
 \Psi_j(\rho_{\mathcal S})
 =\sum_{\lambda,\kappa}
 \Xi_j(x_\lambda,x_\kappa)
 P_\lambda\rho_{\mathcal S}P_\kappa,
 \label{eq:appendix-system-channel}
\end{equation}
with $\Xi_j(x,y)=\frac12\sum_{\sigma\in\{+,-\}}
\chi_j(z_\sigma(x),z_\sigma(y))$.
More explicitly, for an orthonormal controller basis $\{\ket t\}_{t=1}^d$, a reduced-system Kraus family is
\begin{equation}
 L_{j,n,t,\sigma}
 =\frac1{\sqrt2}\sum_\lambda
 \langle t|k_{j,n}(z_\sigma(x_\lambda))\rangle P_\lambda.
 \label{eq:system-Kraus}
\end{equation}
Each operator acts as a scalar on every eigenspace of $X$, including degenerate eigenspaces. 
Its squared coefficients sum to the diagonal $\Xi_j(x_\lambda,x_\lambda)$, giving the system-level terminal effect and accepted spectral law.
The same formulas remain valid in the presence of an untouched reference register upon tensoring each Kraus operator with the reference identity.

For the compiled renewal trial, the clock returns to its initial memory state after every complete trial. 
Thus the controller basis index may be omitted and the accepted Kraus amplitudes are
\begin{equation}
 \ell_{n,\sigma}(x)
 =\frac{g_{\mathsf a}(z_\sigma(x))
 g_{\mathsf c}(z_\sigma(x))^n}{\sqrt2},
 \label{eq:renewal-system-amplitudes}
\end{equation}
which agrees with the construction of Sec.~\ref{subsec:kernel-design}. 
At the walk level, their sum gives
\begin{equation}
 \chi_{\mathsf a}(z,w)=
 \frac{g_{\mathsf a}(z)\overline{g_{\mathsf a}(w)}}
 {1-g_{\mathsf c}(z)\overline{g_{\mathsf c}(w)}}.
 \label{eq:appendix-renewal-kernel}
\end{equation}
The system-level Gram kernel and its correlation-channel factorization then follow from the reduced Kraus family.

\section{Robustness of the rational spectral transformation}
\label{app:robustness}

We give two complementary stability estimates: one for a perturbation of the walk unitary used by a compiled renewal trial, and one for a perturbation of the Hermitian operator inside the target rational function. 
Both estimates make explicit the role of the termination gap or denominator conditioning.

\subsection{Perturbations of the walk oracle}

Let a complete trial contain $L$ controlled-walk queries with branch operators $G_j(W)$, and set
\begin{equation}
 E_j(W)=G_j(W)^\dagger G_j(W),
 \qquad j\in\{\mathsf a,\mathsf r,\mathsf c\}.
 \label{eq:robust-branch-effects}
\end{equation}
The exact compiled trial has the uniform gap
\begin{equation}
 I_{\mathcal W}-E_{\mathsf c}(W)
 \succeq\tau_{\min}I_{\mathcal W},
 \qquad \tau_{\min}=\min_{x\in[-1,1]}S(x)Q(x)>0.
 \label{eq:robust-gap}
\end{equation}

\begin{theorem}[Robustness to walk perturbations]
\label{thm:rus-robustness}
Let $W$ and $\widetilde W$ be unitaries on the same walk space, and apply the \emph{same} compiled trial circuit with $W$ replaced by $\widetilde W$ in every query. 
Suppose
\begin{equation}
 \|\widetilde W-W\|\leq\varepsilon,
 \label{eq:walk-perturbation}
\end{equation}
and
\begin{equation}
 2L\varepsilon<\tau_{\min}.
 \label{eq:robust-small-error}
\end{equation}
Then the perturbed renewal process terminates uniformly and its acceptance effect satisfies
\begin{equation}
 \|\widetilde F_{\mathsf a}-F_{\mathsf a}\|
 \leq\frac{4L\varepsilon}{\tau_{\min}-2L\varepsilon}.
 \label{eq:rus-robustness-bound}
\end{equation}
Consequently, for every input state $\rho_{\mathcal W}$,
\begin{equation}
 \left|\operatorname{Tr}
 [\rho_{\mathcal W}(\widetilde F_{\mathsf a}-F_{\mathsf a})]
 \right|
 \leq\frac{4L\varepsilon}{\tau_{\min}-2L\varepsilon}.
 \label{eq:probability-robustness}
\end{equation}
\end{theorem}

\begin{proof}
A telescoping product bound for the $L$ controlled-walk sections gives
\begin{equation}
 \|\widetilde V-V\|\leq L\varepsilon,
 \qquad \|\widetilde G_j-G_j\|\leq L\varepsilon,
 \label{eq:trial-telescope}
\end{equation}
where $V$ and $\widetilde V$ are the full trial unitaries before branch measurement. 
Each branch operator is a contraction, so
\begin{equation}
 \|\widetilde E_j-E_j\|\leq2L\varepsilon=:\delta.
 \label{eq:effect-perturbation}
\end{equation}
Set $A=I_{\mathcal W}-E_{\mathsf c}$ and
$\widetilde A=I_{\mathcal W}-\widetilde E_{\mathsf c}$. The
unperturbed gap yields
\begin{equation}
 \widetilde A\succeq(\tau_{\min}-\delta)I_{\mathcal W}.
 \label{eq:perturbed-gap}
\end{equation}
Since $\widetilde E_{\mathsf c}=\widetilde G_{\mathsf c}^\dagger
\widetilde G_{\mathsf c}$, this inequality implies
$\|\widetilde G_{\mathsf c}\|^2\leq
1-\tau_{\min}+\delta<1$ and hence uniform termination.

For each of the two walks separately, the branch operators are polynomials in that walk and commute with one another. 
The renewal effects consequently have the resolvent forms
\begin{equation}
 F_{\mathsf a}=E_{\mathsf a}A^{-1},\qquad
 \widetilde F_{\mathsf a}
 =\widetilde E_{\mathsf a}\widetilde A^{-1}.
 \label{eq:robust-resolvents}
\end{equation}
Using the resolvent identity and
$\|E_{\mathsf a}A^{-1}\|=\|F_{\mathsf a}\|\leq1$,
\begin{equation}
 \begin{aligned}
 \|\widetilde F_{\mathsf a}-F_{\mathsf a}\|
 &\leq\|\widetilde E_{\mathsf a}-E_{\mathsf a}\|
 \|\widetilde A^{-1}\|\\
 &\quad+\|E_{\mathsf a}A^{-1}\|
 \|A-\widetilde A\|\|\widetilde A^{-1}\|\\
 &\leq\frac{2\delta}{\tau_{\min}-\delta}.
 \end{aligned}
 \label{eq:robust-proof-bound}
\end{equation}
Substitution of $\delta=2L\varepsilon$ gives the stated bound.
The probability estimate follows from $|\operatorname{Tr}(\rho B)|\leq\|B\|$. \end{proof}

In the perturbative regime $L\varepsilon\ll\tau_{\min}$,
\begin{equation}
 \|\widetilde F_{\mathsf a}-F_{\mathsf a}\|
 =O\left(\frac{L\varepsilon}{\tau_{\min}}\right).
 \label{eq:robust-asymptotic}
\end{equation}
The bound assumes that the same perturbed walk operator is used at every query. 
It does not cover independent adversarial errors at individual queries.

\subsection{Perturbations of the Hermitian Signal}

Write $R=P/Q$ with $Q(x)>0$ on $[-1,1]$, and define
\begin{equation}
 \begin{aligned}
 q_{\min}&=\min_{x\in[-1,1]}Q(x),\\
 \Lambda(p)&=\sum_{k=1}^{\deg p}k|p_k|,
 \qquad p(x)=\sum_kp_kx^k,\\
 P_{\max}&=\max_{x\in[-1,1]}|P(x)|.
 \end{aligned}
 \label{eq:rational-conditioning-data}
\end{equation}
For any Hermitian contractions $X$ and $\widetilde X$,
\begin{equation}
 \|R(\widetilde X)-R(X)\|
 \leq C_R\|\widetilde X-X\|,
 \label{eq:rational-input-robustness}
\end{equation}
with the explicit bound
\begin{equation}
 C_R=\frac{\Lambda(P)}{q_{\min}}
 +\frac{P_{\max}\Lambda(Q)}{q_{\min}^2}.
 \label{eq:rational-conditioning-constant}
\end{equation}
Indeed, the telescoping identity $A^k-B^k=\sum_{m=0}^{k-1}A^{k-1-m}(A-B)B^m$ gives $\|p(A)-p(B)\|\leq\Lambda(p)\|A-B\|$ for Hermitian contractions. 
The inverse identity
\begin{equation}
 \begin{aligned}
 &Q(\widetilde X)^{-1}-Q(X)^{-1}\\
 &\qquad=Q(\widetilde X)^{-1}
 [Q(X)-Q(\widetilde X)]Q(X)^{-1}
 \end{aligned}
 \label{eq:inverse-perturbation}
\end{equation}
and $\|Q(X)^{-1}\|\leq q_{\min}^{-1}$, together with the corresponding estimate for $\widetilde X$, complete the proof.
Thus, sensitivity to perturbations is controlled not only by the rational degree but also by the conditioning of its denominator.

\section{Cost-aware renewal compilation}
\label{app:optimization}

The multiplier $S$ leaves the eventual effect unchanged but changes the polynomial trial length and termination probability. 
For a nonconstant target, the trial length is
\begin{equation}
 L(S)=\max\{\deg(SP),\deg[S(Q-P)],\deg(1-SQ)\},
 \label{eq:length-S}
\end{equation}
with identically zero branches omitted. 
We also put $\tau_{\min}(S)=\min_{[-1,1]}S(x)Q(x)$.
Choosing causal spectral factors without unnecessary powers of $z$ ensures that the maximum factor degree equals $L(S)$: a degree-$l$ probability has a nonzero Fourier coefficient at $z^l$, so any causal spectral factor has exponent width at least $l$, which Fej\'er--Riesz factorization attains.

For a given input state and in the worst case over spectral sectors, the costs are
\begin{equation}
 \begin{aligned}
 C_{\mathrm{run}}(S;\rho_{\mathcal S})
 &=L(S)\int_{-1}^1\frac{d\mu_{\rho_{\mathcal S}}(x)}
 {S(x)Q(x)},\\
 C_{\mathrm{wc}}(S)&=\frac{L(S)}{\tau_{\min}(S)}.
 \end{aligned}
 \label{eq:optimization-cost}
\end{equation}
For a fixed input, division by the $S$-independent acceptance probability $\alpha_R$ gives the expected cost per accepted output.
The restart identity does not require independence of the duration of one run and its terminal outcome: the recursion for expected cost reads $C=C_{\mathrm{run}}+(1-\alpha_R)C$.

To optimize over a bounded polynomial family, fix a maximum length
$L_{\max}$ and a termination floor $t_0>0$.
For each integer $1\leq\ell\leq L_{\max}$, optimize over real
polynomials $S$ satisfying
\begin{equation}
 \begin{gathered}
 t\leq S(x)Q(x)\leq1\qquad(-1\leq x\leq1),\\
 t\geq t_0,\qquad
 \deg(SP),\ \deg[S(Q-P)],\ \deg(1-SQ)\leq\ell,
 \end{gathered}
 \label{eq:convex-feasible}
\end{equation}
where the degree of an identically zero branch is ignored.
For the worst-sector objective, maximizing $t$ minimizes the fixed-length bound $\ell/t$. For the state-aware objective, minimize $\ell\int_{-1}^1[S(x)Q(x)]^{-1}d\mu_{\rho_{\mathcal S}}(x)$.
The degree restrictions are affine constraints on the coefficients of $S$, while the interval positivity requirements admit exact univariate sum-of-squares certificates through the Markov--Luk\'acs representation~\cite{Vaidyanathan1993}. 
The state-aware objective is
convex on the strictly positive feasible set.

For a discrete spectral measure with nodes $x_i$ and weights $w_i$, introduce auxiliary variables $y_i$ and the semidefinite constraints
\begin{equation}
 \begin{pmatrix}S(x_i)Q(x_i)&1\\1&y_i\end{pmatrix}
 \succeq0.
 \label{eq:epigraph}
\end{equation}
Minimizing $\ell\sum_iw_i y_i$ gives a finite semidefinite formulation once the polynomial positivity constraints are expressed as sums of squares. 
For continuous measures, numerical quadrature requires a separate integration-error control.

\begin{proposition}[Optimality within a bounded compiler family]
\label{prop:compiler-opt}
Suppose the admissible family with
$L(S)\leq L_{\max}$ and $\tau_{\min}(S)\geq t_0$ is nonempty. 
If the fixed-length convex subproblems are solved globally, selecting the best candidate by its actual length gives a global minimizer of $C_{\mathrm{run}}$ or $C_{\mathrm{wc}}$ within that family.
\end{proposition}

\begin{proof}
For a fixed length cutoff $\ell$, the coefficient constraints form a closed convex set. Since $Q\geq Q_{\min}>0$ and $0\leq SQ\leq1$, the polynomial $S$ is uniformly bounded.
The degree restriction makes the coefficient space finite dimensional, so the feasible set is compact. 
The objective is continuous under the lower bound $SQ\geq t_0$ and attains its minimum. 
Every admissible multiplier appears in the subproblem with cutoff $\ell=L(S)$. 
The optimal fixed-cutoff objective is no larger than that multiplier's objective; a candidate whose actual length is smaller can only improve its achieved cost. 
Taking the best candidate over finitely many cutoffs therefore proves the claim.
\end{proof}

\section{Resolvent cost and statistical estimation}
\label{app:resolvent-cost}

\subsection{Exact cost and minimax property}

The factors in Eq.~\eqref{eq:res-factors} satisfy $\sum_j|g_j(z)|^2=1$ on $\mathbb T$, and their acceptance branch obeys
\begin{equation}
 G_{\mathsf a}^\dagger G_{\mathsf a}
 =\frac{I_{\mathcal W}}{K+1}.
 \label{eq:res-constant-hazard}
\end{equation}
After any sequence of continuations and fresh-copy restarts, the conditional probability of acceptance in the next one-query trial is therefore $1/(K+1)$. 
The number of queries to the first accepted output is geometric, proving the exact identity $C_{\mathrm{out}}^{\mathrm{RUS}}=K+1$.

The construction is minimax optimal among the polynomial-multiplier renewal compilers considered here. 
To see this, use the normalized resolvent form $R(x)=\eta/(1+x+\eta)$ and evaluate the admissibility condition $S(x)(1+x+\eta)\leq1$ at $x=1$. This gives
\begin{equation}
 p_{\mathsf a}(1)=\eta S(1)
 \leq\frac{\eta}{2+\eta}=\frac1{K+1}.
 \label{eq:minimax-endpoint}
\end{equation}
For an input supported on the top-energy eigenspace, each complete trial uses at least one query and accepts with probability $p_{\mathsf a}(1)$. 
Therefore every such compiler costs at least $K+1$ queries in expectation on that input; the one-query realization attains this value for all inputs. 
This is a minimax statement within the specified renewal family, not a lower bound against arbitrary quantum algorithms.

The expected cost of a single terminal run is
\begin{equation}
 \begin{aligned}
 C_{\mathrm{run}}(\rho_{\mathcal S})
 &=\int_0^\Lambda
 \frac{\Lambda+\gamma}{\lambda+\gamma}
 \,d\mu_{\rho_{\mathcal S}}(\lambda)\\
 &=(K+1)\alpha_\gamma(\rho_{\mathcal S}).
 \end{aligned}
 \label{eq:res-run-cost}
\end{equation}
Division by the terminal acceptance probability gives again $C_{\mathrm{out}}^{\mathrm{RUS}}=K+1$.

The acceptance probability also satisfies the useful soft-window bounds
\begin{equation}
 \begin{aligned}
 \frac12\mu_{\rho_{\mathcal S}}([0,\gamma])
 &\leq\alpha_\gamma(\rho_{\mathcal S})\\
 &\leq\mu_{\rho_{\mathcal S}}([0,\gamma])
 +\gamma\int_{(\gamma,\Lambda]}
 \frac{d\mu_{\rho_{\mathcal S}}(\lambda)}{\lambda}.
 \end{aligned}
 \label{eq:soft-bounds}
\end{equation}
They follow by comparing $\gamma/(\lambda+\gamma)$ with simple upper and lower envelopes below and above $\gamma$.

\subsection{Estimating the resolvent spectral weight}

Repeated independent runs also estimate $a=\alpha_\gamma(\rho_{\mathcal S})$.
If $N$ is the number of terminal runs needed to obtain $M\geq2$ acceptances, then
\begin{equation}
 \Pr(N=n)=\binom{n-1}{M-1}a^M(1-a)^{n-M},
 \qquad n\geq M.
 \label{eq:negative-binomial}
\end{equation}
The estimator $\widehat a=(M-1)/(N-1)$ is unbiased: using
\begin{equation}
 \frac{M-1}{n-1}\binom{n-1}{M-1}
 =\binom{n-2}{M-2},
 \label{eq:unbiased-identity}
\end{equation}
and summing the negative-binomial series gives $\mathbb E\widehat a=a$.
For $0<\varepsilon<1$, standard multiplicative Chernoff bounds applied to the equivalent fixed-number Bernoulli counts yield
relative error at most $\varepsilon$ with failure probability at most $\delta$ once $M=O(\varepsilon^{-2}\log(2/\delta))$.
For $a=1$, the estimate is deterministic.

At the elementary trial level, the acceptance hazard is the constant $1/(K+1)$ even across rejected runs. 
Therefore the number of walk queries required for $M$ acceptances is itself negative binomial, with expectation $M(K+1)$ and high-probability bound
\begin{equation}
 N_W^{(M)}
 =O\left((K+1)[M+\log(1/\delta)]\right)
 \label{eq:estimation-tail}
\end{equation}
with probability at least $1-\delta$.

\section{Explicit factors for the Lorentzian window}
\label{app:lorentz}

Recall the Lorentzian trial in Sec.~\ref{sec:lorentz}, where $R_{\xi,\eta}(x)=\eta^2/[(x-\xi)^2+\eta^2]$, $b=1+|\xi|$, and $M=b^2+\eta^2$.
Set $a_+=b+\xi$ and $a_-=b-\xi$, both at least one. 
For $t\geq1$, define
\begin{equation}
 u_t=\sqrt{\frac{t+\sqrt{t^2-1}}2},
 \quad
 v_t=\sqrt{\frac{t-\sqrt{t^2-1}}2}.
 \label{eq:linear-factor-data}
\end{equation}
These coefficients satisfy $u_t^2+v_t^2=t$ and $2u_tv_t=1$.
Consequently, for $x=(z+z^{-1})/2$ and $z\in\mathbb T$,
\begin{equation}
 |u_t-v_tz|^2=t-x,
 \qquad |u_t+v_tz|^2=t+x.
 \label{eq:linear-factor-identity}
\end{equation}
An explicit causal factorization of the three Lorentzian trial probabilities is
\begin{align}
 g_{\mathsf a}(z)&=\frac{\eta}{\sqrt M},
 \label{eq:L-ga}\\
 g_{\mathsf r}(z)&=\frac{1-2\xi z+z^2}{2\sqrt M},
 \label{eq:L-gr}\\
 g_{\mathsf c}(z)&=
 \frac{(u_{a_+}-v_{a_+}z)(u_{a_-}+v_{a_-}z)}{\sqrt M}.
 \label{eq:L-gc}
\end{align}
Indeed, $g_{\mathsf r}(z)=z[x(z)-\xi]/\sqrt M$, while
\begin{equation}
 \begin{aligned}
 |g_{\mathsf c}(z)|^2
 &=\frac{(b+\xi-x)(b-\xi+x)}{M}\\
 &=\frac{b^2-(x-\xi)^2}{M}.
 \end{aligned}
 \label{eq:L-check}
\end{equation}
Together with $|g_{\mathsf a}|^2=\eta^2/M$, these identities establish the trial normalization $\sum_j|g_j(z)|^2=1$. 
Each factor has degree at most two, so two degree-stripping steps produce the explicit one-trial controlled-walk circuit.

For a centered window $\xi=0$, one has $b=a_+=a_-=1$ and $u_1=v_1=1/\sqrt2$. 
The factors reduce to
\begin{equation}
 (g_{\mathsf a},g_{\mathsf r},g_{\mathsf c})
 =\frac1{\sqrt{1+\eta^2}}
 \left(\eta,\frac{1+z^2}{2},\frac{1-z^2}{2}\right).
 \label{eq:centered-L}
\end{equation}
The accepting probability of each complete trial is the constant $\eta^2/M$. 
A trial uses two controlled-walk queries, giving the exact accepted-output cost $2M/\eta^2$ in Eq.~\eqref{eq:lorentz-cost}. 
Within a terminal run, the one-trial termination probability is $\tau(x)=[(x-\xi)^2+\eta^2]/M$. 
Its minimum over the signal interval is $\tau_{\min}=\eta^2/M$, establishing the geometric tail and expected query bounds used for spectroscopy.

\section{Details of the numerical experiments}
\label{app:numerical_experiments}

We numerically evaluate the first-hitting statistics of the
two RUS filters considered in Sec.~V. Both examples use
a four-dimensional system with an initial state
$|\psi_0\rangle=\sum_{k=1}^{4}\sqrt{w_k}|k\rangle$,
where $|k\rangle$ are eigenvectors of the signal operator.

For the regularized resolvent, the parameters are
\begin{equation}
\begin{gathered}
A=\operatorname{diag}(0,0.05,0.25,0.80),
\quad \Lambda=1,\quad \gamma=0.08,\\
(w_1,w_2,w_3,w_4)=(0.10,0.25,0.30,0.35).
\end{gathered}
\end{equation}
The normalized signal is $X=2A/\Lambda-I$. For the Lorentzian window, we choose
\begin{equation}
\begin{gathered}
X=\operatorname{diag}(-0.82,-0.28,0.12,0.63),\\
\xi=0.10,\quad \eta=0.09,\\
(w_1,w_2,w_3,w_4)=(0.12,0.33,0.40,0.15).
\end{gathered}
\end{equation}

The first-hitting distributions are computed deterministically
by propagating the unnormalized continuation state through
successive complete clock traversals, without Monte Carlo
sampling. The numerical probabilities are verified against
the analytical spectral expressions, with absolute discrepancies
below $10^{-15}$.

The total acceptance probabilities are $0.358392$ and
$0.404046$, and the expected stopping times are $4.838287$
and $60.761535$ complete traversals for the resolvent
and Lorentzian filters, respectively. These stopping times
refer to a single run terminating in either acceptance or
rejection, rather than the total cost per accepted output
including restarts.

\bibliography{bibliography.bib}

@article{Fejer1916,
  author = {Fej{\'e}r, Leopold},
  title = {{\"U}ber trigonometrische Polynome},
  journal = {Journal f{\"u}r die reine und angewandte Mathematik},
  volume = {146},
  pages = {53--82},
  year = {1916},
  doi = {10.1515/crll.1916.146.53}
}

@article{Riesz1916,
  author = {Riesz, Friedrich},
  title = {{\"U}ber ein Problem des Herrn Carath{\'e}odory},
  journal = {Journal f{\"u}r die reine und angewandte Mathematik},
  volume = {146},
  pages = {83--87},
  year = {1916},
  doi = {10.1515/crll.1916.146.83}
}

@article{BerntsonSunderhauf2025,
  author = {Berntson, Bjorn K. and S{\"u}nderhauf, Christoph},
  title = {Complementary Polynomials in Quantum Signal Processing},
  journal = {Communications in Mathematical Physics},
  volume = {406},
  pages = {161},
  year = {2025},
  doi = {10.1007/s00220-025-05302-9}
}

@misc{ChaoEtAl2020,
  author = {Chao, Rui and Ding, Dawei and Gily{\'e}n, Andr{\'a}s and Huang, Cupjin and Szegedy, Mario},
  title = {Finding Angles for Quantum Signal Processing with Machine Precision},
  year = {2020},
  eprint = {2003.02831},
  archivePrefix = {arXiv},
  primaryClass = {quant-ph}
}

@article{Ying2022,
  author = {Ying, Lexing},
  title = {Stable Factorization for Phase Factors of Quantum Signal Processing},
  journal = {Quantum},
  volume = {6},
  pages = {842},
  year = {2022},
  doi = {10.22331/q-2022-10-20-842}
}

@article{NiYing2026,
  author = {Ni, Hongkang and Ying, Lexing},
  title = {Fast Phase Factor Finding for Quantum Signal Processing},
  journal = {SIAM Journal on Scientific Computing},
  volume = {48},
  number = {4},
  pages = {B651--B670},
  year = {2026},
  doi = {10.1137/24M1705214}
}

@article{Schur1917,
  author = {Schur, Issai},
  title = {{\"U}ber Potenzreihen, die im Innern des Einheitskreises beschr{\"a}nkt sind},
  journal = {Journal f{\"u}r die reine und angewandte Mathematik},
  volume = {147},
  pages = {205--232},
  year = {1917},
  doi = {10.1515/crll.1917.147.205}
}

@article{WienerMasani1957,
  author = {Wiener, Norbert and Masani, Pesi},
  title = {The Prediction Theory of Multivariate Stochastic Processes. {I}. The Regularity Condition},
  journal = {Acta Mathematica},
  volume = {98},
  pages = {111--150},
  year = {1957},
  doi = {10.1007/BF02404472}
}

@article{AlexisMnatsakanyanThiele2024,
  author = {Alexis, Maxime and Mnatsakanyan, Gevorg and Thiele, Christoph},
  title = {Quantum Signal Processing and Nonlinear Fourier Analysis},
  journal = {Revista Matem{\'a}tica Complutense},
  volume = {37},
  pages = {655--694},
  year = {2024},
  doi = {10.1007/s13163-024-00494-5}
}

@article{AlexisEtAl2026,
  author = {Alexis, Maxime and Lin, Lin and Mnatsakanyan, Gevorg and Thiele, Christoph and Wang, Jiasu},
  title = {Infinite Quantum Signal Processing for Arbitrary {Szeg\H{o}} Functions},
  journal = {Communications on Pure and Applied Mathematics},
  volume = {79},
  pages = {123--174},
  year = {2026},
  doi = {10.1002/cpa.70007}
}

@article{KikuchiEtAl2023,
  author = {Kikuchi, Yuta and Mc Keever, Conor and Coopmans, Luuk and Lubasch, Michael and Benedetti, Marcello},
  title = {Realization of Quantum Signal Processing on a Noisy Quantum Computer},
  journal = {npj Quantum Information},
  volume = {9},
  pages = {93},
  year = {2023},
  doi = {10.1038/s41534-023-00762-0}
}

@article{MahadevDeWolf2015,
  author  = {Mahadev, Urmila and de Wolf, Ronald},
  title   = {Rational Approximations and Quantum Algorithms with Postselection},
  journal = {Quantum Information and Computation},
  volume  = {15},
  number  = {3--4},
  pages   = {295--307},
  year    = {2015},
  doi     = {10.26421/QIC15.3-4-5},
  eprint  = {1401.0912},
  archivePrefix = {arXiv}
}

@article{Montanaro2016,
  author = {Montanaro, Ashley},
  title = {Quantum Algorithms: An Overview},
  journal = {npj Quantum Information},
  volume = {2},
  pages = {15023},
  year = {2016},
  doi = {10.1038/npjqi.2015.23}
}

@article{Lloyd1996,
  author = {Lloyd, Seth},
  title = {Universal Quantum Simulators},
  journal = {Science},
  volume = {273},
  number = {5278},
  pages = {1073--1078},
  year = {1996},
  doi = {10.1126/science.273.5278.1073}
}

@inproceedings{Szegedy2004,
  author = {Szegedy, Mario},
  title = {Quantum Speed-Up of Markov Chain Based Algorithms},
  booktitle = {Proceedings of the 45th Annual IEEE Symposium on Foundations of Computer Science},
  pages = {32--41},
  year = {2004},
  doi = {10.1109/FOCS.2004.53}
}

@article{PoulinWocjan2009,
  author = {Poulin, David and Wocjan, Pawel},
  title = {Preparing Ground States of Quantum Many-Body Systems on a Quantum Computer},
  journal = {Physical Review Letters},
  volume = {102},
  pages = {130503},
  year = {2009},
  doi = {10.1103/PhysRevLett.102.130503}
}

@article{SapovaFedorov2022,
  author = {Sapova, Mariia D. and Fedorov, Aleksey K.},
  title = {Variational Quantum Eigensolver Techniques for Simulating Carbon Monoxide Oxidation},
  journal = {Communications Physics},
  volume = {5},
  pages = {199},
  year = {2022},
  doi = {10.1038/s42005-022-00982-4}
}

@article{LuchnikovEtAl2022,
  author = {Luchnikov, Ilya A. and Kiktenko, Evgeniy O. and Gavreev, Maxim A. and Ouerdane, Henni and Filippov, Sergey N. and Fedorov, Aleksey K.},
  title = {Probing Non-Markovian Quantum Dynamics with Data-Driven Analysis: Beyond ``Black-Box'' Machine-Learning Models},
  journal = {Physical Review Research},
  volume = {4},
  pages = {043002},
  year = {2022},
  doi = {10.1103/PhysRevResearch.4.043002}
}

@article{ZolotarevEtAl2023,
  author = {Zolotarev, Y. F. and Luchnikov, Ilya A. and L{\'o}pez-Sald{\'i}var, J. A. and Fedorov, Aleksey K. and Kiktenko, Evgeniy O.},
  title = {Continuous Monitoring for Noisy Intermediate-Scale Quantum Processors},
  journal = {Physical Review Applied},
  volume = {19},
  pages = {014027},
  year = {2023},
  doi = {10.1103/PhysRevApplied.19.014027}
}

@article{Gawlik2019,
  author = {Gawlik, Evan S.},
  title = {Zolotarev Iterations for the Matrix Square Root},
  journal = {SIAM Journal on Matrix Analysis and Applications},
  volume = {40},
  number = {2},
  pages = {696--719},
  year = {2019},
  doi = {10.1137/18M1178529}
}

@article{HaleHighamTrefethen2008,
  author = {Hale, Nicholas and Higham, Nicholas J. and Trefethen, Lloyd N.},
  title = {Computing {$A^\alpha$}, {$\log(A)$}, and Related Matrix Functions by Contour Integrals},
  journal = {SIAM Journal on Numerical Analysis},
  volume = {46},
  number = {5},
  pages = {2505--2523},
  year = {2008},
  doi = {10.1137/070700607}
}

@article{GuettelPolizziTangViaud2015,
  author = {Guettel, Stefan and Polizzi, Eric and Tang, Ping Tak Peter and Viaud, Gautier},
  title = {Zolotarev Quadrature Rules and Load Balancing for the {FEAST} Eigensolver},
  journal = {SIAM Journal on Scientific Computing},
  volume = {37},
  number = {4},
  pages = {A2100--A2122},
  year = {2015},
  doi = {10.1137/140980090}
}

@book{Economou2006,
  author = {Economou, Eleftherios N.},
  title = {Green's Functions in Quantum Physics},
  edition = {3},
  series = {Springer Series in Solid-State Sciences},
  volume = {7},
  publisher = {Springer},
  year = {2006},
  doi = {10.1007/3-540-28841-4}
}

@article{TrefethenWeideman2014,
  author = {Trefethen, Lloyd N. and Weideman, J. A. C.},
  title = {The Exponentially Convergent Trapezoidal Rule},
  journal = {SIAM Review},
  volume = {56},
  number = {3},
  pages = {385--458},
  year = {2014},
  doi = {10.1137/130932132}
}

@article{DaviesLewis1970,
  author = {Davies, E. B. and Lewis, J. T.},
  title = {An Operational Approach to Quantum Probability},
  journal = {Communications in Mathematical Physics},
  volume = {17},
  number = {3},
  pages = {239--260},
  year = {1970},
  doi = {10.1007/BF01647093}
}

@book{WisemanMilburn2009,
  author = {Wiseman, Howard M. and Milburn, Gerard J.},
  title = {Quantum Measurement and Control},
  publisher = {Cambridge University Press},
  year = {2009},
  doi = {10.1017/CBO9780511813948}
}

@article{SoedaMurao2021,
  author = {Soeda, Akihito and Murao, Mio},
  title = {Success-or-Draw: A Strategy Allowing Repeat-Until-Success in Quantum Computation},
  journal = {Physical Review Letters},
  volume = {126},
  pages = {150504},
  year = {2021},
  doi = {10.1103/PhysRevLett.126.150504}
}

@article{MizutaFujii2024,
  author = {Mizuta, Kaoru and Fujii, Keisuke},
  title = {Recursive Quantum Eigenvalue and Singular-Value Transformation: Analytic Construction of Matrix Sign Function by Newton Iteration},
  journal = {Physical Review Research},
  volume = {6},
  pages = {L012007},
  year = {2024},
  doi = {10.1103/PhysRevResearch.6.L012007}
}

@article{ChildsKothariSomma2017,
  author = {Childs, Andrew M. and Kothari, Robin and Somma, Rolando D.},
  title = {Quantum Algorithm for Systems of Linear Equations with Exponentially Improved Dependence on Precision},
  journal = {SIAM Journal on Computing},
  volume = {46}, number = {6}, pages = {1920--1950}, year = {2017},
  doi = {10.1137/16M1087072}
}

@article{BerryEtAl2015,
  author = {Berry, Dominic W. and Childs, Andrew M. and Cleve, Richard and Kothari, Robin and Somma, Rolando D.},
  title = {Simulating Hamiltonian Dynamics with a Truncated Taylor Series},
  journal = {Physical Review Letters}, volume = {114}, pages = {090502}, year = {2015},
  doi = {10.1103/PhysRevLett.114.090502}
}

@article{Preskill2018,
  author = {Preskill, John}, title = {Quantum Computing in the {NISQ} Era and Beyond},
  journal = {Quantum}, volume = {2}, pages = {79}, year = {2018},
  doi = {10.22331/q-2018-08-06-79}
}

@article{Roggero2020,
  author = {Roggero, Alessandro},
  title = {Spectral-Density Estimation with the Gaussian Integral Transform},
  journal = {Physical Review A}, volume = {102}, pages = {022409}, year = {2020},
  doi = {10.1103/PhysRevA.102.022409}
}

@book{Watrous2018,
  author = {Watrous, John}, title = {The Theory of Quantum Information},
  publisher = {Cambridge University Press}, year = {2018},
  doi = {10.1017/9781316848142}
}

@article{Feynman1982,
  author  = {Feynman, Richard P.},
  title   = {Simulating Physics with Computers},
  journal = {International Journal of Theoretical Physics},
  volume  = {21},
  pages   = {467--488},
  year    = {1982},
  doi     = {10.1007/BF02650179}
}

@article{Shor1997,
  author  = {Shor, Peter W.},
  title   = {Polynomial-Time Algorithms for Prime Factorization and Discrete Logarithms on a Quantum Computer},
  journal = {SIAM Journal on Computing},
  volume  = {26},
  number  = {5},
  pages   = {1484--1509},
  year    = {1997},
  doi     = {10.1137/S0097539795293172}
}

@inproceedings{Grover1996,
  author    = {Grover, Lov K.},
  title     = {A Fast Quantum Mechanical Algorithm for Database Search},
  booktitle = {Proceedings of the Twenty-Eighth Annual ACM Symposium on Theory of Computing},
  pages     = {212--219},
  year      = {1996},
  publisher = {ACM},
  doi       = {10.1145/237814.237866}
}

@article{HarrowEtAl2009,
  author  = {Harrow, Aram W. and Hassidim, Avinatan and Lloyd, Seth},
  title   = {Quantum Algorithm for Linear Systems of Equations},
  journal = {Physical Review Letters},
  volume  = {103},
  pages   = {150502},
  year    = {2009},
  doi     = {10.1103/PhysRevLett.103.150502}
}

@article{LowChuang2017,
  author  = {Low, Guang Hao and Chuang, Isaac L.},
  title   = {Optimal Hamiltonian Simulation by Quantum Signal Processing},
  journal = {Physical Review Letters},
  volume  = {118},
  pages   = {010501},
  year    = {2017},
  doi     = {10.1103/PhysRevLett.118.010501}
}

@article{LowChuang2019,
  author  = {Low, Guang Hao and Chuang, Isaac L.},
  title   = {Hamiltonian Simulation by Qubitization},
  journal = {Quantum},
  volume  = {3},
  pages   = {163},
  year    = {2019},
  doi     = {10.22331/q-2019-07-12-163}
}

@inproceedings{GilyenEtAl2019,
  author    = {Gily{\'e}n, Andr{\'a}s and Su, Yuan and Low, Guang Hao and Wiebe, Nathan},
  title     = {Quantum Singular Value Transformation and Beyond: Exponential Improvements for Quantum Matrix Arithmetics},
  booktitle = {Proceedings of the 51st Annual ACM SIGACT Symposium on Theory of Computing},
  pages     = {193--204},
  year      = {2019},
  publisher = {ACM},
  doi       = {10.1145/3313276.3316366}
}

@article{MartynEtAl2021,
  author  = {Martyn, John M. and Rossi, Zane M. and Tan, Andrew K. and Chuang, Isaac L.},
  title   = {Grand Unification of Quantum Algorithms},
  journal = {PRX Quantum},
  volume  = {2},
  pages   = {040203},
  year    = {2021},
  doi     = {10.1103/PRXQuantum.2.040203}
}

@article{Haah2019,
  author  = {Haah, Jeongwan},
  title   = {Product Decomposition of Periodic Functions in Quantum Signal Processing},
  journal = {Quantum},
  volume  = {3},
  pages   = {190},
  year    = {2019},
  doi     = {10.22331/q-2019-10-07-190}
}

@article{DongEtAl2021,
  author  = {Dong, Yulong and Meng, Xiang and Whaley, K. Birgitta and Lin, Lin},
  title   = {Efficient Phase-Factor Evaluation in Quantum Signal Processing},
  journal = {Physical Review A},
  volume  = {103},
  pages   = {042419},
  year    = {2021},
  doi     = {10.1103/PhysRevA.103.042419}
}

@article{YoderEtAl2014,
  author  = {Yoder, Theodore J. and Low, Guang Hao and Chuang, Isaac L.},
  title   = {Fixed-Point Quantum Search with an Optimal Number of Queries},
  journal = {Physical Review Letters},
  volume  = {113},
  pages   = {210501},
  year    = {2014},
  doi     = {10.1103/PhysRevLett.113.210501}
}

@article{LinTong2020,
  author  = {Lin, Lin and Tong, Yu},
  title   = {Near-Optimal Ground State Preparation},
  journal = {Quantum},
  volume  = {4},
  pages   = {372},
  year    = {2020},
  doi     = {10.22331/q-2020-12-14-372}
}

@article{DongLinTong2022,
  author  = {Dong, Yulong and Lin, Lin and Tong, Yu},
  title   = {Ground-State Preparation and Energy Estimation on Early Fault-Tolerant Quantum Computers via Quantum Eigenvalue Transformation of Unitary Matrices},
  journal = {PRX Quantum},
  volume  = {3},
  pages   = {040305},
  year    = {2022},
  doi     = {10.1103/PRXQuantum.3.040305}
}

@article{MotlaghWiebe2024,
  author  = {Motlagh, Danial and Wiebe, Nathan},
  title   = {Generalized Quantum Signal Processing},
  journal = {PRX Quantum},
  volume  = {5},
  pages   = {020368},
  year    = {2024},
  doi     = {10.1103/PRXQuantum.5.020368}
}

@article{LuLiuLin2026,
  author  = {Lu, Xi and Liu, Yuan and Lin, Hongwei},
  title   = {Quantum Signal Processing and Quantum Singular Value Transformation on {$U(N)$}},
  journal = {Quantum},
  volume  = {10},
  pages   = {2048},
  year    = {2026},
  doi     = {10.22331/q-2026-03-27-2048}
}

@article{DongEtAl2024Infinite,
  author  = {Dong, Yulong and Lin, Lin and Ni, Hongkang and Wang, Jiasu},
  title   = {Infinite Quantum Signal Processing},
  journal = {Quantum},
  volume  = {8},
  pages   = {1558},
  year    = {2024},
  doi     = {10.22331/q-2024-12-10-1558}
}

@misc{DongAnNiu2024,
  author        = {Dong, Yulong and An, Dong and Niu, Murphy Yuezhen},
  title         = {Feedforward Quantum Singular Value Transformation},
  year          = {2024},
  eprint        = {2408.07803},
  archivePrefix = {arXiv},
  primaryClass  = {quant-ph}
}

@misc{ShenEtAl2024Rational,
  author        = {Shen, Yizhi and Van Buggenhout, Niel and Camps, Daan and Klymko, Katherine and Van Beeumen, Roel},
  title         = {Quantum Rational Transformation Using Linear Combinations of Hamiltonian Simulations},
  year          = {2024},
  eprint        = {2408.07742},
  archivePrefix = {arXiv},
  primaryClass  = {quant-ph}
}

@article{PaetznickSvore2014,
  author  = {Paetznick, Adam and Svore, Krysta M.},
  title   = {Repeat-Until-Success: Non-Deterministic Decomposition of Single-Qubit Unitaries},
  journal = {Quantum Information and Computation},
  volume  = {14},
  number  = {15--16},
  pages   = {1277--1301},
  year    = {2014},
  doi     = {10.26421/QIC14.15-16-2}
}

@article{BocharovEtAl2015,
  author  = {Bocharov, Alex and Roetteler, Martin and Svore, Krysta M.},
  title   = {Efficient Synthesis of Universal Repeat-Until-Success Quantum Circuits},
  journal = {Physical Review Letters},
  volume  = {114},
  pages   = {080502},
  year    = {2015},
  doi     = {10.1103/PhysRevLett.114.080502}
}

@book{Akhiezer1956,
  author    = {Achieser, N. I.},
  title     = {Theory of Approximation},
  publisher = {Frederick Ungar Publishing},
  address   = {New York},
  year      = {1956}
}

@article{Newman1964,
  author  = {Newman, Donald J.},
  title   = {Rational Approximation to $|x|$},
  journal = {Michigan Mathematical Journal},
  volume  = {11},
  number  = {1},
  pages   = {11--14},
  year    = {1964}
}

@book{Higham2008,
  author    = {Higham, Nicholas J.},
  title     = {Functions of Matrices: Theory and Computation},
  publisher = {Society for Industrial and Applied Mathematics},
  address   = {Philadelphia},
  year      = {2008},
  doi       = {10.1137/1.9780898717778}
}

@article{Guettel2013,
  author  = {G{\"u}ttel, Stefan},
  title   = {Rational Krylov Approximation of Matrix Functions: Numerical Methods and Optimal Pole Selection},
  journal = {GAMM-Mitteilungen},
  volume  = {36},
  number  = {1},
  pages   = {8--31},
  year    = {2013},
  doi     = {10.1002/gamm.201310002}
}

@article{Polizzi2009,
  author  = {Polizzi, Eric},
  title   = {Density-Matrix-Based Algorithm for Solving Eigenvalue Problems},
  journal = {Physical Review B},
  volume  = {79},
  pages   = {115112},
  year    = {2009},
  doi     = {10.1103/PhysRevB.79.115112}
}

@article{NakatsukasaEtAl2018,
  author  = {Nakatsukasa, Yuji and S{\`e}te, Olivier and Trefethen, Lloyd N.},
  title   = {The {AAA} Algorithm for Rational Approximation},
  journal = {SIAM Journal on Scientific Computing},
  volume  = {40},
  number  = {3},
  pages   = {A1494--A1522},
  year    = {2018},
  doi     = {10.1137/16M1106122}
}

@article{TrefethenWilber2025,
  author  = {Trefethen, Lloyd N. and Wilber, Heather D.},
  title   = {Computation of Zolotarev Rational Functions},
  journal = {SIAM Journal on Scientific Computing},
  volume  = {47},
  number  = {4},
  pages   = {A2205--A2220},
  year    = {2025},
  doi     = {10.1137/24M1687960}
}

@book{Kailath1980,
  author    = {Kailath, Thomas},
  title     = {Linear Systems},
  publisher = {Prentice-Hall},
  address   = {Englewood Cliffs, NJ},
  year      = {1980}
}

@book{Alpay2001,
  author    = {Alpay, Daniel},
  title     = {The Schur Algorithm, Reproducing Kernel Spaces and System Theory},
  publisher = {American Mathematical Society},
  address   = {Providence, RI},
  year      = {2001},
  isbn      = {978-0-8218-2155-8}
}

@book{Vaidyanathan1993,
  author    = {Vaidyanathan, P. P.},
  title     = {Multirate Systems and Filter Banks},
  publisher = {Prentice Hall},
  address   = {Englewood Cliffs, NJ},
  year      = {1993}
}

@book{Dumitrescu2007,
  author    = {Dumitrescu, Bogdan},
  title     = {Positive Trigonometric Polynomials and Signal Processing Applications},
  publisher = {Springer},
  address   = {Dordrecht},
  year      = {2007},
  doi       = {10.1007/978-1-4020-5125-8}
}

@article{BallBolotnikov2004,
  author  = {Ball, Joseph A. and Bolotnikov, Vladimir},
  title   = {Realization and interpolation for {Schur--Agler}-class functions on domains with matrix polynomial defining function in {$\mathbb{C}^n$}},
  journal = {Journal of Functional Analysis},
  volume  = {213},
  number  = {1},
  pages   = {45--87},
  year    = {2004},
  doi     = {10.1016/j.jfa.2004.04.008},
  url     = {https://doi.org/10.1016/j.jfa.2004.04.008}
}

@inproceedings{AlaouiMahoney2015,
  author    = {El Alaoui, Ahmed and Mahoney, Michael W.},
  title     = {Fast Randomized Kernel Ridge Regression with Statistical Guarantees},
  booktitle = {Advances in Neural Information Processing Systems 28},
  pages     = {775--783},
  year      = {2015},
  publisher = {Curran Associates, Inc.},
  url       = {https://proceedings.neurips.cc/paper/2015/hash/f3f27a324736617f20abbf2ffd806f6d-Abstract.html},
  eprint    = {1411.0306},
  archivePrefix = {arXiv},
  primaryClass  = {stat.ML}
}

@inproceedings{MuscoMusco2017,
  author    = {Musco, Cameron and Musco, Christopher},
  title     = {Recursive Sampling for the {Nystr{\"o}m} Method},
  booktitle = {Advances in Neural Information Processing Systems 30},
  pages     = {3834--3846},
  year      = {2017},
  publisher = {Curran Associates, Inc.},
  url       = {https://proceedings.neurips.cc/paper/2017/hash/a03fa30821986dff10fc66647c84c9c3-Abstract.html},
  eprint    = {1605.07583},
  archivePrefix = {arXiv},
  primaryClass  = {cs.LG}
}

@article{Rall2020,
  author  = {Rall, Patrick},
  title   = {Quantum algorithms for estimating physical quantities using block encodings},
  journal = {Physical Review A},
  volume  = {102},
  number  = {2},
  pages   = {022408},
  year    = {2020},
  doi     = {10.1103/PhysRevA.102.022408},
  url     = {https://doi.org/10.1103/PhysRevA.102.022408}
}

@article{EfrosEtAl2007,
  author  = {Efros, Victor D. and Leidemann, Winfried and Orlandini, Giuseppina and Barnea, Nir},
  title   = {The Lorentz integral transform ({LIT}) method and its applications to perturbation-induced reactions},
  journal = {Journal of Physics G: Nuclear and Particle Physics},
  volume  = {34},
  number  = {12},
  pages   = {R459--R528},
  year    = {2007},
  doi     = {10.1088/0954-3899/34/12/R02},
  url     = {https://doi.org/10.1088/0954-3899/34/12/R02},
  eprint  = {0708.2803},
  archivePrefix = {arXiv},
  primaryClass  = {nucl-th}
}

@incollection{DritschelRovnyak2010,
  author    = {Dritschel, Michael A. and Rovnyak, James},
  title     = {The Operator {Fej{\'e}r--Riesz} Theorem},
  booktitle = {A Glimpse at Hilbert Space Operators: Paul R. Halmos in Memoriam},
  editor    = {Axler, Sheldon and Rosenthal, Peter and Sarason, Donald},
  series    = {Operator Theory: Advances and Applications},
  volume    = {207},
  pages     = {223--254},
  publisher = {Birkh{\"a}user},
  address   = {Basel},
  year      = {2010},
  doi       = {10.1007/978-3-0346-0347-8_14},
  url       = {https://doi.org/10.1007/978-3-0346-0347-8_14},
  eprint    = {0903.3639},
  archivePrefix = {arXiv},
  primaryClass  = {math.FA}
}

@article{Kalman1963,
  author  = {Kalman, Rudolf E.},
  title   = {Mathematical Description of Linear Dynamical Systems},
  journal = {Journal of the Society for Industrial and Applied Mathematics, Series A: Control},
  volume  = {1},
  number  = {2},
  pages   = {152--192},
  year    = {1963},
  doi     = {10.1137/0301010},
  url     = {https://doi.org/10.1137/0301010}
}

@book{ZhouDoyleGlover1996,
  author    = {Zhou, Kemin and Doyle, John C. and Glover, Keith},
  title     = {Robust and Optimal Control},
  publisher = {Prentice Hall},
  address   = {Upper Saddle River, NJ},
  year      = {1996},
  isbn      = {0-13-456567-3}
}

@article{PlonkaTasche2026,
  author  = {Plonka, Gerlind and Tasche, Manfred},
  title   = {Fast computation of best approximation polynomials for the reciprocal function},
  journal = {Numerical Algorithms},
  year    = {2026},
  doi     = {10.1007/s11075-026-02364-1},
  url     = {https://doi.org/10.1007/s11075-026-02364-1}
}

@misc{fedorov2022quantum,
  title={Quantum computing at the quantum advantage threshold: a down-to-business review},
  author={Fedorov, A K and Gisin, N and Beloussov, S M and Lvovsky, A I},
  archivePrefix = {arXiv},
  primaryClass = {quant-ph},
  year={2022},
  eprint = {2203.17181}
}

\end{document}